\documentclass[a4paper,11pt]{article}
\usepackage{jheppub} 
\usepackage[T1]{fontenc} 
\usepackage{physics}
\RequirePackage{verbatim}
\NewDocumentCommand{\tens}{t_}
{
	\IfBooleanTF{#1}
	{\tensop}
	{\otimes}%
}
\NewDocumentCommand{\tensop}{m}
{
	\mathbin{\mathop{\otimes}\displaylimits_{#1}}%
}
\usepackage{graphicx}
\usepackage{wrapfig}
\usepackage{amsmath}
\usepackage{amsfonts}
\usepackage{amssymb}
\usepackage{color}
\usepackage{subcaption}
\usepackage{dsfont}
\usepackage{mathrsfs}
\usepackage{dcolumn}
\usepackage{bm}
\usepackage{hyperref}
\usepackage{verbatim}
\usepackage{float}
\usepackage[title]{appendix}

\title{\textbf{\boldmath Single and entangled atomic systems in thermal bath and the Fulling-Davies-Unruh effect}}

\author[a,1]{Arnab Mukherjee\note{Corresponding author.},}
\author[b]{Sunandan Gangopadhyay,}
\author[c]{Archan. S. Majumdar}

\affiliation{\textit{ Department of Astrophysics and High Energy Physics, S.N.~Bose National Centre for Basic Sciences, JD Block, Sector-III, Salt Lake, Kolkata 700106, India}}
\emailAdd{arnab.mukherjee@bose.res.in}
\emailAdd{sunandan.gangopadhyay@gmail.com}
\emailAdd{archan@bose.res.in}

\abstract{\noindent In this study, we revisit the Fulling-Davies-Unruh effect in the context of two-level single and entangled atomic systems that are static in a thermal bath. We consider the interaction between the systems and a massless scalar field, covering the scenarios of free space as well as within a cavity. Through the calculation of atomic transition rates and comparing with the results of [\textcolor{blue}{\textit{Phys. Rev. D 108 (2023) 085018}}], it is found that in free space there is an equivalence between the upward and downward transition rates of an uniformly accelerated atom with respect to an observer with that of a single atom which is static with respect to the observer and immersed in a thermal bath, as long as the temperature of the thermal bath matches the Unruh temperature. This equivalence  between the upward and downward transition rates breaks down in the presence of a cavity. For two-atom systems, considering the initial state to be in a general pure entangled form, we find that in this case the equivalence between the upward and downward transition rates of the accelerated and static thermal bath scenarios  holds only under specific limiting conditions in free space, but breaks down completely in a cavity setup. Though the ratio of the upward and downward transition rates in the thermal bath matches exactly with those of the accelerated systems in free space as well as inside the cavity.}

\begin{document} 
	\maketitle
	\flushbottom

\section{Introduction}\label{sec:Intro}

The physics of interaction of two-level atomic systems with quantum fields is endowed with rich and diverse properties \cite{Plenio, Amico, Ficek2003, Ficek2004,  Crispino1, Hongwei1, Zhu3, Zhu2, Lin3, Zhu1, Zhou3, Zhu}. 
Investigations into the radiative properties of a uniformly accelerated single atom \cite{Hongwei1, Zhu3, Zhu2, Lin3, Zhu1, Zhou3} have been extended to scenarios involving multiple atoms interacting with the massless scalar field and the electromagnetic field \cite{Passante4, Passante3, Passante2, Passante1, Passante, Menezes1, Menezes, Menezes2, Zhou2, Lima2019}.
A two-level atom serves as a detector, and coupled to an external field, can give rise to the phenomenon of the Fulling-Davies-Unruh (FDU) effect \cite{Fulling, Davies1975, Unruh, birrell1984quantum}. Emanating from the study of quantum field theory in curved background, the FDU effect reveals that a uniformly accelerated atom feels a thermal bath in the Minkowski vacuum, with the temperature $T$ being related to the proper acceleration $\alpha$ as $T=\alpha/2\pi$. The FDU effect  has a deep connection with black hole thermodynamics and the information loss paradox \cite{Unruh, Crispino, Ross, Matsas1}. 

It has been understood that the upward and downward transition rates of the accelerated detector, which in this case is the atom, is exactly the same as seen by a local inertial observer and by an observer who is coaccelerating with the detector. This theoretical equivalence between the transition rates has been found to hold for a single atom in free space as well as in the setting of a single reflecting boundary \cite{Zhu1}.  A similar equivalence also holds at the level of transition rates for an entangled atom in free space \cite{Zhou2020} provided a thermal bath at the FDU temperature exists in the coaccelerated frame. Moreover, it has been recently shown \cite{Mukherjee2023} that the equivalence between the upward and downward transition rates in the context of an uniformly accelerating Unruh-DeWitt detector and a coaccelerating detector immersed in  a thermal bath holds completely in free space as well as inside a cavity, for both the single and entangled atomic systems. However, it has also been found \cite{Zhou2020} that for an entangled atom in free space this theoretical equivalence between the transition rates breaks down completely when the coaccelerated frame scenario is replaced with a static atom immersed in a background thermal bath with temperature  equal to the Unruh temperature.

The primary motivation of the present work is to establish further the status of equivalence stemming from the FDU effect. Though the equivalence dictated by the FDU effect is universally valid at the conceptual level, its revelation in real physical or experimental scenarios is contingent on the choice of set-ups or contexts. Context plays a very important role in manifestation of quantum features. Quantum contextuality is a well-studied subject with foundational implications (see, \cite{Contextuality_review_2022} for a recent review) as well as diverse applications \cite{Gupta_2023}, and entanglement is known to elucidate certain subtle features of contextuality. In this work we consider entangled atomic systems as a detector model to study whether the equivalence is  manifested for such set-ups involving entanglement.

Moreover, currently there exists a wide upsurge of interest in quantum entanglement in relativistic settings, intertwining profound concepts from quantum field theory, information theory and gravitational physics \cite{Frolov, Valentini1991, OLevin, Svaiter1, Matsas3, Muller1, Muller, Haroche, Passante5, Matsas2, Alsing2003, Alsing2004, Fuentes, Alsing, Bermudez, Hwang1, Hwang, Richter}.  Initially uncorrelated Unruh-DeWitt detectors may get entangled by interacting locally with the vacuum state of a quantum field \cite{Reznik2003}. Localized detectors can extract non-local correlations from the  a quantum field through the process of entanglement harvesting  \cite{Salton2015, Martinez2015, Martinez2016, Martinez2016_1, Henderson2018, Henderson2020, Henderson2021}. Relativistic quantum information explores how entanglement is affected by not only Lorentz boosts \cite{Chatterjee2017}, but by non-inertial effects as well \cite{Terno2004, Chatterjee2022}.

In practical scenarios, the degradation of entanglement due to uncontrolled coupling with external fields is a genuine  concern, and extensive research has been undertaken to investigate the transition rates between the states of entangled atoms moving in various trajectories, yielding a rich paradigm of possibilities \cite{Massar, Franson, Lin, Lin1, Svaiter2, Hu2015, Cai2019, Picanco2020, Zhou2020a}. Since configurations in analogue cavity QED such as superconducting circuits \cite{PhysRevB.92.064501} and laser-driven technologies \cite{Bulanov2006, Eichmann2009, McWilliams2012} can achieve substantial accelerations, such systems are beginning to be used towards experimental evidence of theoretical results in relativistic quantum information. Reflecting boundaries have been shown to play a significant role  in such studies on relativistic quantum phenomena in superconducting circuits \cite{PhysRevLett.110.113602, PhysRevB.92.064501}, as well as in  secure quantum communication over long distances \cite{huang2019protection, huang2020deterministic, PhysRevD.104.105020, Aberg_2013}.

Due to the apparently contradictory nature regarding the status of equivalence between the transition rates as manifested from certain findings of \cite{Zhu1, Zhou2020, Mukherjee2023}, it becomes pertinent to explore further the implications of replacing the coaccelerated frame scenario with a static frame immersed in a background thermal bath. The aim of our present study is a comprehensive investigation of the FDU phenomenon by computing the upward and downward transition rates in the context of a static Unruh-DeWitt detector immersed in a background thermal bath and comparing with those of an accelerating Unruh-DeWitt detector given in \cite{Mukherjee2023}. Our study pertains to single two-level atoms as well as an entangled two-atom systems either in free space, and confined in a cavity. For the case of two-atom systems, one of our main focus is on the role played by quantum entanglement on the transition rates in the presence of boundaries, which in turn, have a direct bearing on revelation of the FDU effect, as shown through our subsequent analysis.

From a fundamental perspective, the impact of cavity setup on atom-field interactions and radiative processes of entangled atoms are manifold  \cite{Scully_2003, Scully_2006, haroche2006exploring, zhang_2007, zhou2013boundary, liu2021entanglement, arias2016boundary, Cheng_2018, Zhang2019}. It was observed  \cite{Zhou2, Passante, Chatterjee2021_2} that  reflecting boundaries strongly influence the resonance interaction energy of uniformly accelerated two-atom system. In \cite{Chatterjee2021_1}, it was found that reflecting boundaries induce effects which lead to the violation of equivalence in an accelerating atom-mirror system in the generalized uncertainty principle framework. Reflecting boundaries have several other
interesting consequences in the context of quantum entanglement \cite{zhang_2007, zhou2013boundary, Cheng_2018, liu2021entanglement}, and quantum thermodynamics \cite{Mukherjee2022}. This line of inquiry presents a possibility for advancing our understanding about the FDU effect within a cavity quantum electrodynamics (QED) framework, that is important both from fundamental and practical points of view.  

The paper is organised as follows. In section \ref{sec:Int_th}, we  analyze the transition rates for the cases when a single and two entangled static atomic systems interact with a massless scalar field in a background thermal bath in empty space and in the presence of a cavity, respectively. In section \ref{sec:Ex}, we calculate a quantitative estimation of the upward transition rate of a single and two atom system inside a cavity by considering the Rubidium and Caesium atom. We  conclude with a summary of our results in section \ref{sec:Con}. Throughout the paper, we take $\hbar=c=k_B=1$, where $k_B$ is the Boltzmann constant.
\section{Interaction of the static atomic system with a thermal bath}\label{sec:Int_th}
In this section, we start by investigating the case when a static atomic system interacts with a massless scalar field in a thermal state at an arbitrary temperature $T$ and undergoes transitions in between its lower and higher energy states.

\subsection{Single atom system}\label{sec:Trate_thS}
In this analysis, we also consider a single atom (an Unruh-DeWitt detector) with two energy levels $\vert g\rangle$ and $\vert e\rangle$ with corresponding energy values $-\omega_{0}/2$ and $+\omega_{0}/2$, remains static in a thermal state of a massless scalar field.
Following the procedure given in the subsection (\ref{sec:CoupS}), the transition probability from the initial state $\vert i\rangle$ to the final state $\vert f\rangle$ can be written as
\begin{equation}
\mathscr{P}^{\beta}_{\vert i\rangle\rightarrow \vert f\rangle}=\lambda^2 \vert m_{fi}\vert^2 \mathcal{F}^{\beta}(\Delta E)\,\label{transprobS}
\end{equation}
where $\Delta E= E_{f}-E_{i}$, $m_{fi}=\langle f\vert m(0)\vert i\rangle$ and
the response function $\mathcal{F}^{\beta}(\Delta E)$ is defined as 
\begin{equation}
\mathcal{F}^{\beta}(\Delta E)=\int_{-\infty}^{+\infty} d\tau \int_{-\infty}^{+\infty} d\tau'\,e^{-i\Delta E(\tau-\tau')}\,G^{+}_{\beta}(x(\tau),x(\tau'))\label{responseS}
\end{equation}
where  
\begin{equation}
G^{+}_{\beta}(x(\tau),x(\tau'))=\frac{tr[e^{-\beta \mathscr{H}_{F}}\phi(x(\tau))\phi(x(\tau')]}{tr[e^{-\beta \mathscr{H}_{F}}]}\label{wightmanS}
\end{equation}
is the positive frequency Wightman function of the massless scalar field in a thermal state at an arbitrary temperature $T$ with $\mathscr{H}_F=\displaystyle\sum_{\mathbf{k}}\omega_\mathbf{k} a^{\dagger}_\mathbf{k} a_\mathbf{k}$ \cite{birrell1984quantum}.

Exploiting the time translational invariance property of the positive frequency Wightman function, the response function per unit proper time can be written as
\begin{equation}
\mathscr{F}^{\beta}(\Delta E)=\int_{-\infty}^{+\infty} d(\Delta\tau) \,e^{-i\Delta E\Delta\tau}\,G^{+}_{\beta}(x(\tau),x(\tau'))\label{responserateS}
\end{equation}
where $\Delta\tau=\tau-\tau'$.
Therefore, the transition probability per unit proper time from the initial state $\vert i\rangle$ to the final state $\vert f\rangle$ turns out to be
\begin{equation}
\mathscr{R}^{\beta}_{\vert i\rangle\rightarrow \vert f \rangle}=\lambda^2 \vert m_{fi}\vert^2 \mathscr{F}^{\beta}(\Delta E)\,.\label{transprobrateS}
\end{equation}

\noindent In the next subsections, we employ the above equations to study the transitions of a static single atom interacting with a thermal state of a massless scalar field in both empty space and a cavity.
\subsubsection{Transition rates for single atom system in empty space}
We initially take into account the transition rates of a single atom that is interacting with a massless thermal scalar field in the empty space. In the laboratory frame, atomic trajectory is given by
\begin{equation}
t=\tau,\,\,\,x=y=z=0\,\label{trajecS}
\end{equation}
where $\tau$ denote the proper time of the atom.

The thermal Wightman function is given by \cite{Zhou2020} (see Appendix \ref{Appendix:BB} for a detailed calculation)
\begin{align}
&G^{+}_{\beta}(x(\tau),x(\tau'))\nonumber\\
=&-\frac{1}{4\pi^2}\displaystyle\sum_{n=-\infty}^{\infty}\frac{1}{(t(\tau)-t(\tau')-in\beta-i\varepsilon)^2-(x(\tau)-x(\tau'))^2-(y(\tau)-y(\tau'))^2-(z(\tau)-z(\tau'))^2}\,.\label{wightmn1S}
\end{align}
Substituting \eqref{trajecS} in \eqref{wightmn1S}, the thermal Wightman function turns out to be \cite{birrell1984quantum}
\begin{equation}
 G^{+}_{\beta}(x(\tau),x(\tau'))=-\frac{1}{4\pi^2}\displaystyle\sum_{n=-\infty}^{\infty}\frac{1}{(\Delta\tau-in\beta-i\varepsilon)^2}\,. \label{Wght1S}
\end{equation}
Substituting the Wightman function into eq.\eqref{responserateS} and eq.\eqref{transprobrateS}, the transition rate from the initial state $\vert i\rangle$ to the final state $\vert f\rangle$ becomes 
\begin{equation}
\mathscr{R}^{\beta}_{\vert i\rangle\rightarrow \vert f\rangle}=-\frac{\lambda^2\vert m_{fi}\vert^2}{4\pi^2}\displaystyle\sum_{n=-\infty}^{\infty}\int_{-\infty}^{+\infty} d(\Delta\tau) \,e^{-i\Delta E\Delta\tau}\frac{1}{(\Delta\tau-in\beta-i\varepsilon)^2}\,. \label{transprobrate1S}
\end{equation}
 Simplifying the transition rates, eq.\eqref{transprobrate1S}, by performing the contour integration \cite{freitag2009complex} as shown in Appendix A of \cite{Mukherjee2023}, we obtain
 \begin{equation}
\mathscr{R}^{\beta}_{\vert i\rangle\rightarrow \vert f\rangle}=\frac{\lambda^2\vert m_{fi}\vert^2\vert \Delta E\vert}{2\pi}\left[\theta\,(-\Delta E)\left(1+\frac{1}{\exp(\vert \Delta E\vert/T)-1}\right)+\theta\,(\Delta E)\left(\frac{1}{\exp(\Delta E/T)-1}\right)\right] \label{transrateES}
\end{equation}
where $\theta(\Delta E)$ is the Heaviside step function defined as
\begin{equation}\label{delE}
\theta(\Delta E)=
\begin{cases}
1,\,\,\,\Delta E>0\\
0,\,\,\,\Delta E<0.
\end{cases}
\end{equation}
Eq. \eqref{transrateES} reveals that the two transition processes, namely, the upward and downward transition can take place even when the atom is static but placed inside a thermal bath. From the above equation, it may also be noted that the upward transition or excitation process is solely depends on the temperature of the thermal bath. Considering the initial state $\vert i\rangle=\vert g\rangle$, final state $\vert f\rangle=\vert e\rangle$ and vice-versa and using the definition $m_{eg}=\langle e\vert m(0)\vert g\rangle$, we obtain $\vert m_{ge}\vert^2=\vert m_{eg}\vert^2=1$, and $\Delta E=\omega_0$ for the transition $g\rightarrow e$ and $\Delta E=-\,\omega_0$ for the transition $e\rightarrow g$. Using the above results the upward and downward transition rate takes the form 
\begin{equation}\label{up_empS}
\mathscr{R}^{\beta}_{\vert g\rangle\rightarrow \vert e\rangle}=\frac{\lambda^2\omega_0}{2\pi}\left(\frac{1}{\exp(\omega_0/T)-1}\right)
\end{equation}
\begin{equation}\label{dwn_empS}
\mathscr{R}^{\beta}_{\vert e\rangle\rightarrow \vert g\rangle}=\frac{\lambda^2\omega_0}{2\pi}\left(1+\frac{1}{\exp(\omega_0/T)-1}\right)\,.
\end{equation} 
Taking the ratio of the above two results, we get
\begin{equation}\label{Ratio}
\frac{\mathscr{R}^{\beta}_{\vert g\rangle\rightarrow \vert e\rangle}}{\mathscr{R}^{\beta}_{\vert e\rangle\rightarrow \vert g\rangle}}\equiv\frac{\mathscr{R}^{\beta}_{up}}{\mathscr{R}^{\beta}_{down}}=\exp(-\omega_0/T)\,.
\end{equation}
\noindent Eq. \eqref{Ratio} is the consequence of a universal relation called  the principle of detailed balance \cite{Crispino2}.
From the above expressions two points can be noted. First, it is seen that the upward transition rate entirely depends on the temperature of the thermal state of the massless scalar field. At $T=0$, the upward transition rate vanishes. Secondly, 
if we take the thermal bath temperature in the static frame at $T=\alpha/2\pi$, then eqs.(\ref{up_empS}), (\ref{dwn_empS}), (\ref{Up_empS}) and  (\ref{Down_empS})
clearly show that the upward and the downward transition rates of an uniformly accelerated atom seen by an instantaneously
inertial observer (see Appendix \ref{Appendix:AA}) and by a static observer in a thermal bath are identical. Further, from eq.\eqref{Ratio} and eq.\eqref{Ratio0}, it can be seen that the ratio in both cases also matches in the limit $T=\alpha/2\pi$. Therefore, for a single atom system in empty space the equivalence between the effect of uniform acceleration and the effect of thermal bath holds at the level of transition rates as well as their ratios.
\subsubsection{Transition rates for single atom system in a cavity}
\begin{figure}[H]
\centering
\includegraphics[scale=0.42]{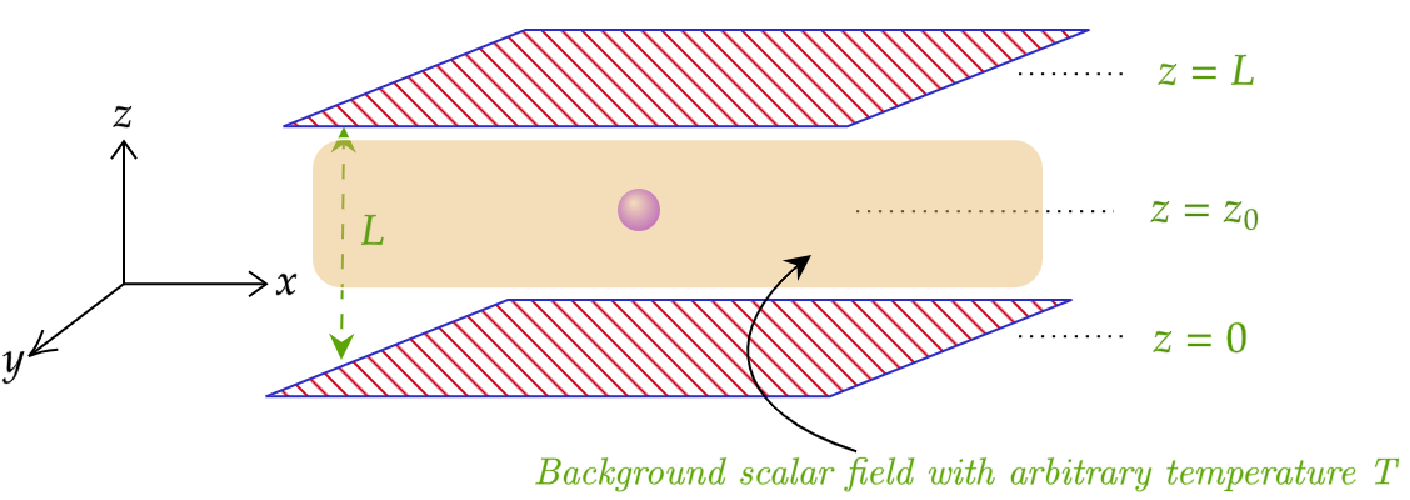}
\caption{Static atom confined in a cavity with a thermal bath at a temperature $T$.}\label{fig:SingleC1_new}
\end{figure}

We now consider that the a static atom is interacting with a massless thermal scalar field inside a cavity of length $L$ as shown in Figure \ref{fig:SingleC1_new}. Assuming the scalar field obeys the Dirichlet boundary condition $ \phi\vert_{z=0}=\phi\vert_{z=L}=0$, the Wightman function of the thermal scalar field confined in the cavity of length $L$ takes the form \cite{Brown_1969} (see Appendix \ref{Appendix:CC} for a detailed calculation)
\begin{align}
&G^{+}_{\beta}(x(\tau),x(\tau'))\nonumber\\
=&-\frac{1}{4\pi^2}\displaystyle\sum_{m=-\infty}^{\infty}\displaystyle\sum_{n=-\infty}^{\infty}\left[\frac{1}{(t(\tau)-(\tau')-im\beta-i\varepsilon)^2-\vert\Delta\mathbf{x}_{\perp}\vert^2-(z(\tau)-z(\tau')-2nL)^2}\right.\nonumber\\
&\left.-\frac{1}{(t(\tau)-t(\tau')-im\beta-i\varepsilon)^2-\vert\Delta\mathbf{x}_{\perp}\vert^2-(z(\tau)+z(\tau')-2nL)^2}\right] \label{WightCS}
\end{align}
with $\vert\Delta\mathbf{x}_{\perp}\vert^2=\sqrt{(x(\tau)-x(\tau'))^2+(y(\tau)-y(\tau'))^2}$.\\
\noindent Inside the cavity the atomic trajectory is given by
\begin{equation}
t(\tau)=\tau,\,\,\,x=y=0,\,\,z=z_{0}\,\label{trajecCS}
\end{equation}
Using the above trajectories in eq.\eqref{WightCS}, the Wightman function becomes
\begin{equation}
 G^{+}_{\beta}(x (\tau),x (\tau'))=-\frac{1}{4\pi^2}
\displaystyle\sum_{m=-\infty}^{\infty}\displaystyle\sum_{n=-\infty}^{\infty}\left[\frac{1}{(\Delta\tau-im\beta-i\varepsilon)^2-\mathfrak{d}_{1}^2}-\frac{1}{(\Delta\tau-im\beta-i\varepsilon)^2-\mathfrak{d}_{2}^2}\right] \label{WghtCS1}
\end{equation}
with $\mathfrak{d}_1=2nL,\,\mathfrak{d}_2=2z_{0}-2nL$.
Using the Wightman function, eq. (\ref{WghtCS1}) into eq.\eqref{responserateS}, the transition rate from the initial state $\vert i\rangle$ to the final state $\vert f\rangle$ is given by
\begin{align}
\mathscr{R}^{\beta}_{\vert i\rangle\rightarrow \vert f\rangle}=&-\frac{\lambda^2\vert m_{fi}\vert^2}{4\pi^2}\displaystyle\sum_{m=-\infty}^{\infty}\displaystyle\sum_{n=-\infty}^{\infty}\left[\int_{-\infty}^{+\infty} d(\Delta\tau) \,e^{-i\Delta E\Delta\tau}\frac{1}{(\Delta\tau-im\beta-i\varepsilon)^2-\mathfrak{d}_{1}^2}\right.\,\nonumber\\
&-\left.\int_{-\infty}^{+\infty} d(\Delta\tau) \,e^{-i\Delta E\Delta\tau}\frac{1}{(\Delta\tau-im\beta-i\varepsilon)^2-\mathfrak{d}_{2}^2}\right]\,. \label{transprobrateC1S}
\end{align}
Simplifying the above equation by following the method of contour integral, rate of transition from the initial state $\vert i\rangle$ to the final state $\vert f\rangle$ can be written as
\begin{align}
\mathscr{R}^{\beta}_{\vert i\rangle\rightarrow \vert f\rangle}=&\lambda^2\vert m_{fi}\vert^2\left[\theta\,(-\Delta E)\left\{\frac{\vert\Delta E\vert}{2\pi}+\mathfrak{q}\left(\vert \Delta E\vert,\,2L\right)-\mathfrak{r}\left(\vert \Delta E\vert,\,2z_{0},2L\right)\right\}\left(1+\frac{1}{\exp(\Delta E/T)-1}\right)\right.\nonumber\\
&+\left.\theta\,(\Delta E)\left\{\frac{\Delta E}{2\pi}+\mathfrak{q}\left(\Delta E,\,2L\right)-\mathfrak{r}\left(\Delta E,\,2z_{0},2L\right)\right\}\left(\frac{1}{\exp(\Delta E/T)-1}\right)\right]\label{transrateCES}
\end{align}
where we have defined
\begin{align}
\mathfrak{q}\left(\Delta E,\,2L\right)&=2\displaystyle\sum_{n=1}^{\infty}\mathfrak{p}\left(\Delta E,\,2nL\right)\label{qfunc1S}\\
\mathfrak{r}\left(\Delta E,\,2z_0,\,2L\right)&=\displaystyle\sum_{n=-\infty}^{\infty}\mathfrak{p}\left(\Delta E,\,2(z_0-nL)\right)\,\label{rfunc1S}
\end{align}
with $\mathfrak{p}\left(\Delta E,\,z_0\right)$ is given by
\begin{equation}
\mathfrak{p}\left(\Delta E,\,z_{0}\right)=\frac{\sin(\Delta E z_{0})}{2\pi z_0}\,.\label{pfunc1S}
\end{equation}
Hence, from the above result the upward and downward transition rates can be written as
\begin{equation}\label{up_CS}
\mathscr{R}^{\beta}_{\vert g\rangle\rightarrow \vert e\rangle}=\lambda^2\left[\left\{\frac{\omega_0}{2\pi}+\mathfrak{q}\left(\omega_0,\,2L\right)-\mathfrak{r}\left(\omega_0,\,2z_{0},2L\right)\right\}\left(\frac{1}{\exp(\omega_0/T)-1}\right)\right]
\end{equation}
\begin{equation}\label{down_CS}
\mathscr{R}^{\beta}_{\vert e\rangle\rightarrow \vert g\rangle}=\lambda^2\left[\left\{\frac{\omega_0}{2\pi}+\mathfrak{q}\left(\omega_0,\,2L\right)-\mathfrak{r}\left(\omega_0,\,2z_{0},2L\right)\right\}\left(1+\frac{1}{\exp(\omega_0/T)-1}\right)\right]\,.
\end{equation}
\noindent From the above analysis it follows that the transitions observed by an instantaneously
inertial observer and a static observer in a thermal
bath for both the upward and the downward transition rates
when the atom is confined in a cavity are clearly distinct. 
We also observe that taking the thermal bath temperature in the static frame $T=\alpha/2\pi$, eqs. (\ref{up_CS}), (\ref{down_CS}), (\ref{Up_CS}) and (\ref{Down_CS}) indicate that there is a non-equivalence between the transition rates of a uniformly accelerated
atom seen by an instantaneously inertial observer and a static atom seen by a static observer in a thermal bath inside the cavity. Though the upward and the downward transition rates of a static atom in a thermal bath are not same as those corresponding to a uniformly accelerated atom,
 it can be seen from eqs. (\ref{up_CS}), (\ref{down_CS}), (\ref{Up_CS}) and (\ref{Down_CS}), that at $T=\alpha/2\pi$, the ratio of the upward and the downward transition rates of a static atom in a thermal bath is identical with that of a uniformly accelerated atom  (eq.\eqref{Ratio1}). From the above analysis, it is also observed that the ratio of eqs. (\ref{up_CS}), (\ref{down_CS}) is identical with the free space result (eq.\eqref{Ratio}).
This is an universal feature independent of the detector model and follows from the principle of detailed balance \cite{Crispino2}.

In order to describe the single boundary and free space scenarios, we now derive the limiting cases of these expressions. Taking the limit $L\rightarrow\infty$, we find that in eq.(s)(\ref{up_CS}, \ref{down_CS}) only $n=0$ term  survives from the infinite summation  and one can effectively reduce the cavity scenario to a situation where only one reflecting boundary exists. Hence, using this limit, the upward and downward transition rates in the presence of a single reflecting boundary turn out to be 
\begin{equation}
\mathscr{R}^{\beta}_{\vert g\rangle\rightarrow \vert e\rangle}=\lambda^2\left[\left\{\frac{\omega_0}{2\pi}-\mathfrak{p}\left(\omega_0,\,2z_{0}\right)\right\}\left(\frac{1}{\exp(\omega_0/T)-1}\right)\right]
\end{equation}
\begin{equation}
\mathscr{R}^{\beta}_{\vert e\rangle\rightarrow \vert g\rangle}=\lambda^2\left[\left\{\frac{\omega_0}{2\pi}-\mathfrak{p}\left(\omega_0,\,2z_{0}\right)\right\}\left(1+\frac{1}{\exp(\omega_0/T)-1}\right)\right]\,.
\end{equation}
On the other hand, taking the limits $L\rightarrow\infty$ and $z_{0}\rightarrow\infty$ together, eq.(s)(\ref{up_CS}, \ref{down_CS}) lead to the expression for the upward and downward transition rates in the free space given by eq.(s)(\ref{up_empS}, \ref{dwn_empS}).

We study the variation of the transition rate of a single two-level atom confined to a cavity, where the parameters are the atom's distance from the boundary ($z_0$), the cavity's length ($L$), and the temperature of the thermal field ($T$). The atom's ground state energy level is $\vert g\rangle$, and its excited state energy level is $\vert e\rangle$. The findings are plotted below, where all physical quantities are expressed in dimensionless units. To fix the dimensionless parameters, we consider a single $^{87}Rb$ atom and take the atomic data from Ref. \cite{Steck2024Rubidium}. Recent time, it is observed that experimentally atomic excitations in nanoscale waveguides \cite{corzo2019waveguide} can be achievable through some novel nanofabrication techniques \cite{Vetsch2010, Goban2012}. Following the Refs. \cite{Steck2024Rubidium, Vetsch2010, Vylegzhanin_2023}, we choose $\omega_0=1.59$ eV and $L=400$ nm. The cavity effect becomes prominent when the all the parameters are comparable \cite{Donaire}. Due to this reason, we take $T$ and $z_0$ in such a way so that $\omega_0L$, $\omega_0 z_0$, and $T/\omega_0$ becomes comparable.

\begin{figure}[H]
\centering
\includegraphics[scale=0.95]{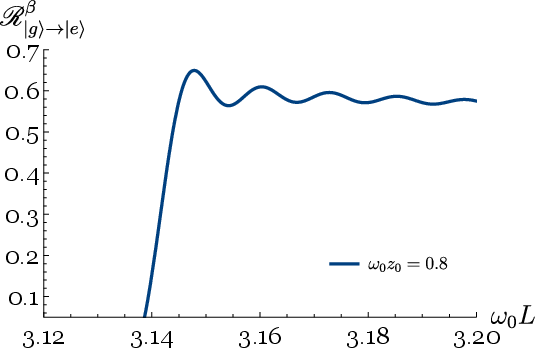}
\caption{Transition rate from $\vert g\rangle\rightarrow\vert e\rangle$ (per unit $\frac{\lambda^2\omega_0}{2\pi}$) versus separation between the two boundaries for a fixed value of $T/\omega_0=1$ and $\omega_0z_0=0.8$.}\label{fig:Single_L}
\end{figure}

\noindent Figure \ref{fig:Single_L} shows the variation of the transition rate from $\vert g\rangle\rightarrow\vert e\rangle$ (per unit $\frac{\lambda^2\omega_0}{2\pi}$) with respect to the cavity length for a fixed value of the distance of the atom from one boundary and the temperature of the thermal field. From the figure, it can be seen that for a fixed value of the initial atomic distance $z_0$ from one boundary, the transition rate initially very low due to the cavity effect and get enhanced after a certain cavity length and get saturated for large values of $L$ ($\omega_0 L>>\omega_0 z_0$). 
This is to be expected as extending the cavity length results in an increased number of field modes participating in the interaction between the atom and the scalar field, which raises the transition rate. When $\omega_0 L>>\omega_0 z_0$, the cavity scenario reduces to the case of a single boundary, and hence, the upward transition rate saturates for large $L$.

\begin{figure}
\centering
\includegraphics[scale=0.95]{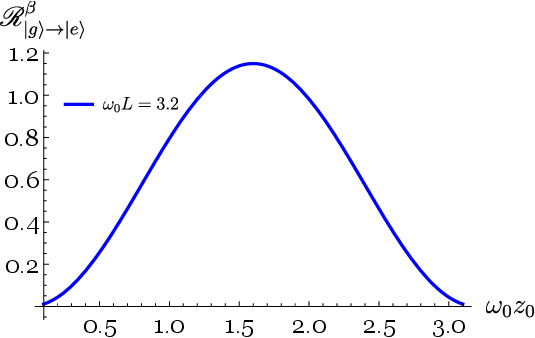}
\caption{Transition rate from $\vert g\rangle\rightarrow\vert e\rangle$ (per unit $\frac{\lambda^2\omega_0}{2\pi}$) versus distance of the atom from one boundary for a fixed value of $T/\omega_0=1$ and $\omega_0L=3.2$.}\label{fig:Single_z}
\end{figure}

\noindent Figure \ref{fig:Single_z} shows the variation of the transition rate from $\vert g\rangle\rightarrow\vert e\rangle$ (per unit $\frac{\lambda^2\omega_0}{2\pi}$) with respect to the distance of the atom from one boundary for a fixed value of the length of the cavity and the temperature of the thermal field. From the figure, it is observed that for a fixed value of the cavity length $L$, when we increase the atomic distance from one boundary, the transition rate increases and at a certain value of $z_0$ it attains a maximum value and then it gets reduced by further increment of $z_0$. The reason behind this is the following. Increasing the atomic distance from the boundary reduces the boundary effect on the number of field modes taking part in the interaction between the atom and the scalar field, which in turn increases the transition rate. This result is consistent with Figure \ref{fig:Single_L}.

\begin{figure}[H]
\centering
\includegraphics[scale=0.95]{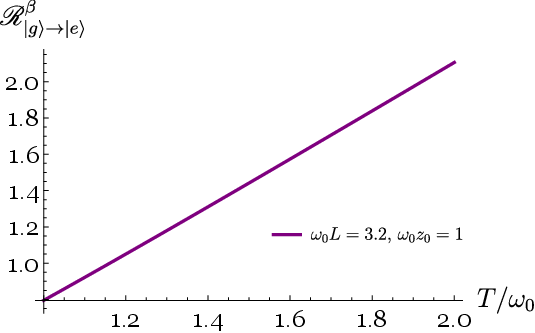}
\caption{Transition rate from $\vert g\rangle\rightarrow\vert e\rangle$ (per unit $\frac{\lambda^2\omega_0}{2\pi}$) versus temperature for a fixed value of $\omega_0 L=3.2$ and $\omega_0z_0=1$.}\label{fig:Single_T}
\end{figure}

\noindent Figure \ref{fig:Single_T} shows the variation of the transition rate from $\vert g\rangle\rightarrow\vert e\rangle$ (per unit $\frac{\lambda^2\omega_0}{2\pi}$) with respect to the temperature of the thermal bath for a fixed value of the cavity length and distance of the atom from one boundary. From the figure, it is observed that for a fixed value of the length of the cavity $L$ and the atomic distance $z_0$ from one boundary, the transition rate increases when the temperature of the thermal bath is increased.
\subsection{Two-atom system}\label{sec:Trate_th}
In this subsection, we analyse the transition rates of a static two-atom system prepared in any generic pure entangled state $\vert\psi\rangle$ that interacts with the massless scalar field in a thermal state at an arbitrary temperature $T$. Following similar procedures as given in subsection \ref{sec:CoupT}, the transition rate of a static two-atom system takes the form
\begin{equation}
\mathscr{R}^{\beta}_{\vert \psi\rangle\rightarrow \vert E_n\rangle}=\lambda^2 \Big[\vert m^{(A)}_{E_n\psi}\vert^2 \mathscr{F}^{\beta}_{AA}(\Delta E)+m^{(B)}_{E_n\psi}\,m^{(A)\,*}_{E_n\psi}\mathscr{F}^{\beta}_{AB}(\Delta E)\Big]+A\rightleftharpoons B\,\text{terms}\label{transprobrateMM}
\end{equation}
where
\begin{equation}
\mathscr{F}^{\beta}_{\xi\xi'}(\Delta E)=\int_{-\infty}^{+\infty} d(\Delta\tau) \,e^{-i\Delta E\Delta\tau}\,G^{+}_{\beta}(x_{\xi}(\tau),x_{\xi'}(\tau'))\label{responserate-M}
\end{equation}
with $\xi,\xi'$ can be labeled by $A$ or $B$, and 
\begin{equation}
G^{+}_{\beta}(x_{\xi}(\tau),x_{\xi'}(\tau'))=\frac{tr[e^{-\beta \mathscr{H}_F}\phi(x_{\xi}(\tau))\phi(x_{\xi'}(\tau')]}{tr[e^{-\beta \mathscr{H}_F}]}\label{wightmanM}
\end{equation}
is the positive frequency Wightman function of the scalar field in a thermal state at an arbitrary temperature $T$, and $\mathscr{H}_F=\displaystyle\sum_{\mathbf{k}}\omega_\mathbf{k} a^{\dagger}_\mathbf{k} a_\mathbf{k}$.
\subsubsection{Transition rates for two-atom system in empty space}
We  take into account the transition rates of a stationary two-atom system which interacts with the massless scalar field in a thermal state at an arbitrary temperature $T$ in the empty space. In the laboratory frame the trajectories of both the atoms read
\begin{equation}
t_{A/B}(\tau)=\tau,\,\,\,x_{A/B}=0,\,\,\,y_{A/B}=0,\,\,z_{A}=0,\,\,z_{B}=d\,.\label{trajecM}
\end{equation}
Using the usual mode expansion of the scalar field operator in eq. \eqref{wightmanM}, Wightman function becomes \cite{Zhou2020}
\begin{align}
&G^{+}_{\beta}(x_{\xi}(\tau),x_{\xi'}(\tau'))\nonumber\\
=&-\frac{1}{4\pi^2}\displaystyle\sum_{m=-\infty}^{\infty}\frac{1}{(t_{\xi}-t'_{\xi'}-im\beta-i\varepsilon)^2-(x_{\xi}-x'_{\xi'})^2-(y_{\xi}-y'_{\xi'})^2-(z_{\xi}-z'_{\xi'})^2}\,.\label{wightmnM1}
\end{align}
Using eq.(s)(\ref{trajecM}, \ref{wightmnM1}) the Wightman function turns out to be
\begin{equation}
 G^{+}_{\beta}(x_{\xi}(\tau),x_{\xi'}(\tau'))=-\frac{1}{4\pi^2}\displaystyle\sum_{m=-\infty}^{\infty}\left[\frac{1}{(\Delta\tau-im\beta-i\varepsilon)^2}\right] \label{WghtM1}
\end{equation}
with $\Delta\tau=\tau-\tau'$ for $\xi=\xi'$, and 
 \begin{equation}
 G^{+}_{\beta}(x_{\xi}(\tau),x_{\xi'}(\tau'))=-\frac{1}{4\pi^2}\displaystyle\sum_{m=-\infty}^{\infty}\left[\frac{1}{(\Delta\tau-im\beta-i\varepsilon)^2-d^2}\right] \label{WghtM2}
\end{equation}
 for $\xi\neq\xi'$.
 
Substituting the Wightman functions, eq.(s) (\ref{WghtM1}, \ref{WghtM2}) into eq.\eqref{transprobrate}, the transition rate of the two-atom system
from the initial state $\vert \psi\rangle$ to the final state $\vert E_n\rangle$ can be rewritten as
\begin{align}
\mathscr{R}^{\beta}_{\vert \psi\rangle\rightarrow \vert E_n\rangle}=\lambda^2 \Big[&\vert m^{(A)}_{E_n\psi}\vert^2 \mathscr{F}^{\beta}_{AA}(\Delta E)+\vert m^{(B)}_{E_n\psi}\vert^2 \mathscr{F}^{\beta}_{BB}(\Delta E)+m^{(B)}_{E_n\psi}\,m^{(A)\,*}_{E_n\psi}\mathscr{F}^{\beta}_{AB}(\Delta E)\nonumber\\
&+m^{(A)}_{E_n\psi}\,m^{(B)\,*}_{E_n\psi}\mathscr{F}^{\beta}_{BA}(\Delta E)\Big] \label{transprobrateM1}
\end{align}
 with
\begin{equation}
\mathscr{F}^{\beta}_{\xi\xi'}(\Delta E)=
 -\frac{1}{4\pi^2}\displaystyle\sum_{m=-\infty}^{\infty}\int_{-\infty}^{+\infty} d(\Delta\tau) \,e^{-i\Delta E\Delta\tau}\left[\frac{1}{(\Delta\tau-im\beta-i\varepsilon)^2}\right]
\end{equation}
for $\xi=\xi'$ and 
\begin{equation}
\mathscr{F}^{\beta}_{\xi\xi'}(\Delta E)=
 -\frac{1}{4\pi^2}\displaystyle\sum_{m=-\infty}^{\infty}\int_{-\infty}^{+\infty} d(\Delta\tau) \,e^{-i\Delta E\Delta\tau}\left[\frac{1}{(\Delta\tau-im\beta-i\varepsilon)^2-d^2}\right]
\end{equation} 
for $\xi\neq\xi'$.

We may simplify the transition rates, eq.\eqref{transprobrateM1}, by performing the above integrations using the contour integration technique 
 \begin{align}
\mathscr{R}^{\beta}_{\vert \psi\rangle\rightarrow \vert E_n\rangle}=&\lambda^2\left\{\theta\,(-\Delta E)\left(\frac{\vert \Delta E\vert}{2\pi}+\frac{\sin 2\theta \sin(\vert \Delta E\vert d)}{2\pi d}\right)\left(1+\frac{1}{\exp{(\vert \Delta E\vert/T)}-1}\right)\right.\nonumber\\
&+\left.\theta\,(\Delta E)\left(\frac{\Delta E}{2\pi}+\frac{\sin 2\theta \sin(\Delta E d)}{2\pi d}\right)\left(\frac{1}{\exp{(\Delta E/T)}-1}\right)\right\}\,. \label{transrateME}
\end{align}
The above equation reveals that the two transition processes, namely, the downward and the upward transition can take place for the two-atom system with the upward transition rate
\begin{equation}\label{upth_emp}
\mathscr{R}^{\beta}_{\vert \psi\rangle\rightarrow \vert e_{A}e_{B}\rangle}=\lambda^2\left\{\left(\frac{\omega_{0}}{2\pi}+\frac{\sin 2\theta \sin(\omega_{0} d)}{2\pi d}\right)\left(\frac{1}{\exp{(\omega_{0}/T)}-1}\right)\right\}\,
\end{equation}
and the downward transition rate
\begin{equation}\label{dwnth_emp}
\mathscr{R}^{\beta}_{\vert \psi\rangle\rightarrow \vert g_{A}g_{B}\rangle}=\lambda^2\left\{\left(\frac{\omega_{0}}{2\pi}+\frac{\sin 2\theta \sin(\omega_{0} d)}{2\pi d}\right)\left(1+\frac{1}{\exp{(\omega_{0}/T)}-1}\right)\right\}\,.
\end{equation}

\noindent From the above analysis it is observed that if we compare the transition rates, eq.(s)(\ref{upth_emp}, \ref{dwnth_emp}), with those of the uniformly accelerated two-atom system as seen by an instantaneously inertial observer, eq.(s)(\ref{up_emp}, \ref{dwn_emp}), we find that the transition rates of the static two-atom system immersed in a thermal bath as seen by a static observer are in general distinct from those of the two-atom system uniformly accelerated in the Minkowski vacuum even when the temperature of the thermal bath is taken to be the FDU temperature  \cite{Zhou2020}. Although, here we would like to point out an additional feature. It is observed that in the limit $\alpha d<<1$, expanding eqs.(\ref{up_emp}) and (\ref{dwn_emp}) and keeping terms upto $\mathcal{O}(\alpha^2 d^2)$ gives the results eqs.(\ref{up_empA}) and (\ref{dwn_empA}). Now from these equations, it is seen that the leading term of the transition rates of uniformly accelerated two-atom system seen by a inertial observer in free space (eqs.(\ref{up_empA}) and (\ref{dwn_empA})) matches with those of the static two-atom system seen by a static observer in thermal bath in free space (eqs.(\ref{upth_emp}) and  (\ref{dwnth_emp})) if we take the temperature of the thermal bath equal to the FDU temperature. This observation implies that in empty space there is an approximate equivalence between the upward and downward transition rates of the scenarios when two static atoms are placed in a thermal bath and two atoms are accelerating uniformly with respect to an inertial observer.

\subsubsection{Transition rates for two-atom system in a cavity}
\begin{figure}[!htbp]
\centering
\includegraphics[scale=0.42]{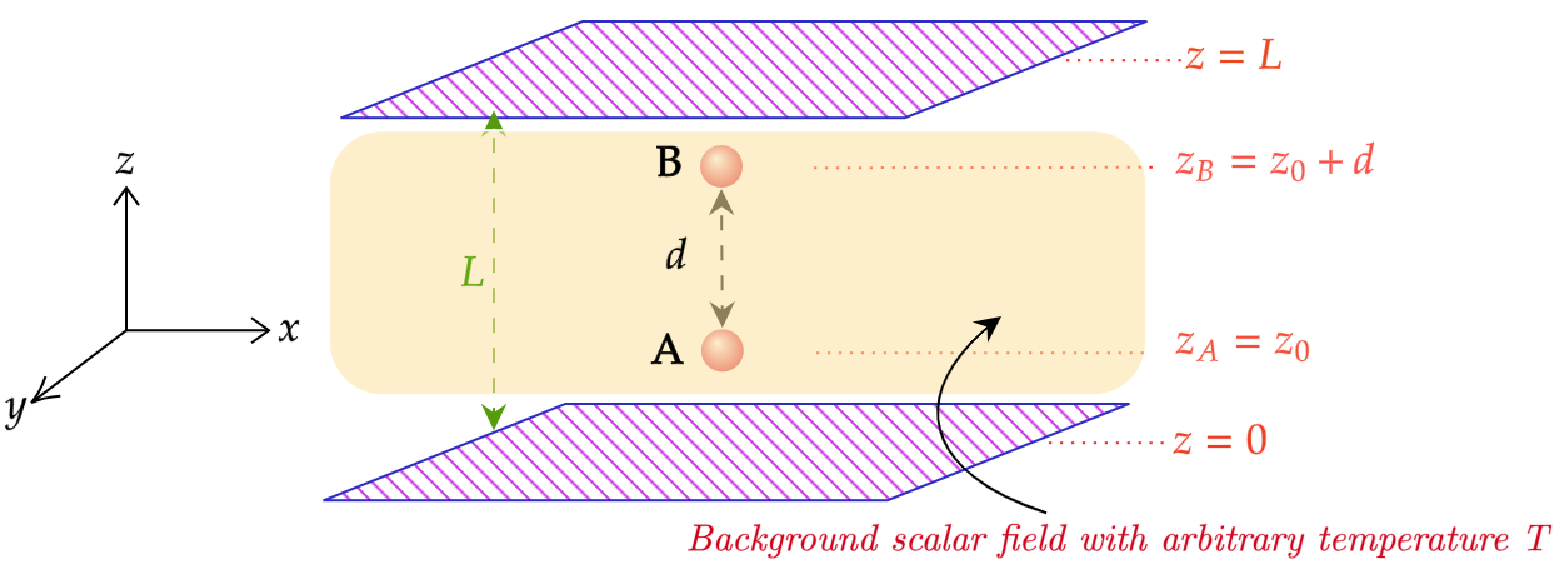}
\caption{Static two-atom confined in a cavity with a thermal bath at a temperature $T$.}\label{fig:DoubleC1_new}
\end{figure}

Let us consider a static two-atom system  interacting with a thermal state of a massless scalar field confined in a cavity of length $L$ as shown in Figure \ref{fig:DoubleC1_new}.
Assuming that the scalar field obeys the Dirichlet boundary condition $ \phi\vert_{z=0}=\phi\vert_{z=L}=0$,  the thermal Wightman function inside the cavity takes the form \cite{Brown_1969}
\begin{align}
&G^{+}_{\beta}(x_{\xi}(\tau),x_{\xi'}(\tau'))\nonumber\\
=&-\frac{1}{4\pi^2}\displaystyle\sum_{m=-\infty}^{\infty}\displaystyle\sum_{n=-\infty}^{\infty}\left[\frac{1}{(t_{\xi}(\tau)-t_{\xi'}(\tau')-im\beta-i\varepsilon)^2-\vert\Delta\mathbf{x}_{\perp}\vert^2-(z_{\xi}(\tau)-z_{\xi'}(\tau')-2nL)^2}\right.\nonumber\\
&\left.-\frac{1}{(t_{\xi}(\tau)-t_{\xi'}(\tau')-im\beta-i\varepsilon)^2-\vert\Delta\mathbf{x}_{\perp}\vert^2-(z_{\xi}(\tau)+z_{\xi'}(\tau')-2nL)^2}\right] \label{WightCM}
\end{align}
with $\vert\Delta\mathbf{x}_{\perp}\vert^2=\sqrt{(x_{\xi}(\tau)-x_{\xi'}(\tau'))^2+(y_{\xi}(\tau)-y_{\xi'}(\tau'))^2}$.\\ 

\noindent In case of two atoms inside the cavity the atomic trajectories take the form
\begin{equation}
t_{A/B}(\tau)=\tau,\,\,\,x_{A/B}=0,\,\,\,y_{A/B}=0,\,\,z_{A}=z_{0},\,\,z_{B}=z_{0}+d\,.\label{trajecCM}
\end{equation}
Using above trajectories in eq.\eqref{WightCM}, the Wightman function becomes
\begin{equation}
 G^{+}_{\beta}(x_{\xi}(\tau),x_{\xi'}(\tau'))=-\frac{1}{4\pi^2}
\displaystyle\sum_{m=-\infty}^{\infty}\displaystyle\sum_{n=-\infty}^{\infty}\left[\frac{1}{(\Delta\tau-im\beta-i\varepsilon)^2-\mathfrak{d'}_{1}^2}-\frac{1}{(\Delta\tau-im\beta-i\varepsilon)^2-\mathfrak{d'}_{2}^2}\right] \label{WghtCM1}
\end{equation}
for $\xi=\xi'$, with $\mathfrak{d'}_1=2nL,\,\mathfrak{d'}_2=2z_{\xi}-2nL$ and 
\begin{equation}
 G^{+}_{\beta}(x_{\xi}(\tau),x_{\xi'}(\tau'))=-\frac{1}{4\pi^2}\displaystyle\sum_{m=-\infty}^{\infty}\displaystyle\sum_{n=-\infty}^{\infty}\left[\frac{1}{(\Delta\tau-im\beta-i\varepsilon)^2-\mathfrak{d'}_{3}^2}-\frac{1}{(\Delta\tau-im\beta-i\varepsilon)^2-\mathfrak{d'}_{4}^2}\right] \label{WghtCM2}
\end{equation}
 for $\xi\neq\xi'$, with $\mathfrak{d'}_3=d+2nL$ (for $\xi=A,\xi'=B$), $\mathfrak{d'}_3=d-2nL$ (for $\xi=B,\xi'=A$) and $\mathfrak{d'}_4=2z_{0}+d-2nL$.

Using above Wightman functions, the rate of transition from the initial entangled state $\vert \psi\rangle$ to the final separable state $\vert E_n\rangle$ can be written as
\begin{align}
\mathscr{R}^{\beta}_{\vert \psi\rangle\rightarrow \vert E_n\rangle}=\lambda^2 \displaystyle\sum_{n=-\infty}^{\infty}\Big[&\vert m^{(A)}_{E_n\psi}\vert^2 \mathscr{F}^{\beta}_{AA}(\Delta E)+\vert m^{(B)}_{E_n\psi}\vert^2 \mathscr{F}^{\beta}_{BB}(\Delta E)+m^{(B)}_{E_n\psi}\,m^{(A)\,*}_{E_n\psi}\mathscr{F}^{\beta}_{AB}(\Delta E)\nonumber\\
&+m^{(A)}_{E_n\psi}\,m^{(B)\,*}_{E_n\psi}\mathscr{F}^{\beta}_{BA}(\Delta E)\Big] \label{transprobrateCM}
\end{align}
 with
\begin{equation}
\mathscr{F}^{\beta}_{\xi\xi'}(\Delta E)= -\frac{1}{4\pi^2}\displaystyle\sum_{m=-\infty}^{\infty}\int_{-\infty}^{+\infty} d(\Delta\tau) \,e^{-i\Delta E\Delta\tau}\left[\frac{1}{(\Delta\tau-im\beta-i\varepsilon)^2-\mathfrak{d'}_{1}^2}-\frac{1}{(\Delta\tau-im\beta-i\varepsilon)^2-\mathfrak{d'}_{2}^2}\right]
\end{equation}
for $\xi=\xi'$ and 
\begin{equation}
\mathscr{F}^{\beta}_{\xi\xi'}(\Delta E)= -\frac{1}{4\pi^2}\displaystyle\sum_{m=-\infty}^{\infty}\int_{-\infty}^{+\infty} d(\Delta\tau) \,e^{-i\Delta E\Delta\tau}\left[\frac{1}{(\Delta\tau-im\beta-i\varepsilon)^2-\mathfrak{d'}_{3}^2}-\frac{1}{(\Delta\tau-im\beta-i\varepsilon)^2-\mathfrak{d'}_{4}^2}\right]
\end{equation} 
for $\xi\neq\xi'$.

Eq.\eqref{transprobrateCM} can be further simplified by performing the above integrations using the contour integration procedure
\begin{align}
\mathscr{R}^{\beta}_{\vert \psi\rangle\rightarrow \vert E_n\rangle}=&\lambda^2\Bigg\{\theta\,(-\Delta E)\left(\frac{\vert \Delta E\vert}{2\pi}+\mathfrak{q}\left(\vert \Delta E\vert,\,2L\right)-\cos^2\theta\,\mathfrak{r}\left(\vert \Delta E\vert,\,2z_{0},2L\right)-\sin^2\theta\mathfrak{s}\left(\vert \Delta E\vert,\,2z_{0},2d,2L\right)\,\right.\Bigg.\nonumber\\
&+\bigg.\left.\sin2\theta\,\mathfrak{t}\left(\vert \Delta E\vert,\,d,2L\right)-\sin2\theta\,\mathfrak{s}\left(\vert \Delta E\vert,\,2z_{0},d,2L\right)\bigg)\left(1+\frac{1}{\exp{(\vert \Delta E\vert/T)}-1}\right)\right.\nonumber\\
&+\theta\,(\Delta E)\left(\frac{\Delta E}{2\pi}+\mathfrak{q}\left(\Delta E,\,2L\right)-\cos^2\theta\,\mathfrak{r}\left(\Delta E,\,2z_{0},2L\right)-\sin^2\theta\mathfrak{s}\left(\Delta E,\,2z_{0},2d,2L\right)\,\right.\nonumber\\
&+\bigg.\left.\sin2\theta\,\mathfrak{t}\left(\Delta E,\,d,2L\right)-\sin2\theta\,\mathfrak{s}\left(\Delta E,\,2z_{0},d,2L\right)\bigg)\left(\frac{1}{\exp{(\Delta E/T)}-1}\right)\right\} \label{transrateCEM}
\end{align}
where we have defined
\begin{align}
\mathfrak{s}\left(\Delta E,\,2z_0\,,d,\,2L\right)&=\displaystyle\sum_{n=-\infty}^{\infty}\mathfrak{p}\left(\Delta E,\,2z_0+d-2nL\right)\,\label{sfuncM}\\
\mathfrak{t}\left(\Delta E,\,d,\,2L\right)&=\displaystyle\sum_{n=-\infty}^{\infty}\mathfrak{p}\left(\Delta E,\,d-2nL\right)\label{tfuncM}
\end{align}
and $\mathfrak{q}\left(\Delta E,\,2L\right),\,
\mathfrak{r}\left(\Delta E,\,2z_0,\,2L\right),\,\mathfrak{p}\left(\Delta E,\,2z_0\right)$ are given in eq.(s)(\ref{qfunc1S}, \ref{rfunc1S}, \ref{pfunc1S}).

Similar to the previous case, the above equation also suggests that two transition process can take place for the two-atom system in presence of a reflecting boundary with the upward transition rate
\begin{align}\label{upth_C1}
\mathscr{R}^{\beta}_{\vert \psi\rangle\rightarrow \vert e_{A}e_{B}\rangle}=&\lambda^2\left\{\bigg(\frac{\omega_0}{2\pi}+\mathfrak{q}\left(\omega_{0},\,2L\right)-\cos^2\theta\,\mathfrak{r}\left(\omega_{0},\,2z_{0},\,2L\right)-\sin^2\theta\,\mathfrak{s}\left(\omega_{0},\,2z_{0},\,2d,\,2L\right)\,\bigg.\right.\nonumber\\
&+\bigg.\left.\sin2\theta\,\mathfrak{t}\left(\omega_{0},\,d,\,2L\right)-\sin2\theta\,\mathfrak{s}\left(\omega_{0},\,2z_{0},\,d,\,2L\right)\bigg)\left(\frac{1}{\exp{(\omega_{0}/T)}-1}\right)\right\}\,
\end{align}
and the downward transition rate
\begin{align}\label{dwnth_C1}
\mathscr{R}^{\beta}_{\vert \psi\rangle\rightarrow \vert g_{A}g_{B}\rangle}=&\lambda^2\left\{\bigg(\frac{\omega_0}{2\pi}+\mathfrak{q}\left(\omega_{0},\,2L\right)-\cos^2\theta\,\mathfrak{r}\left(\omega_{0},\,2z_{0},\,2L\right)-\sin^2\theta\,\mathfrak{s}\left(\omega_{0},\,2z_{0},\,2d,\,2L\right)\,\bigg.\right.\nonumber\\
&+\bigg.\left.\sin2\theta\,\mathfrak{t}\left(\omega_{0},\,d,\,2L\right)-\sin2\theta\,\mathfrak{s}\left(\omega_{0},\,2z_{0},\,d,\,2L\right)\bigg)\left(1+\frac{1}{\exp{(\omega_{0}/T)}-1}\right)\right\}\,.
\end{align}
From the above analysis, eqs (\ref{upth_C1}), (\ref{dwnth_C1}), (\ref{up_C}) and (\ref{dwn_C}) clearly display that transition rates of a uniformly accelerated
two-atom seen by an instantaneously inertial observer and a static two-atom seen by a static observer in a thermal bath are non-identical inside the cavity even if we consider the temperature of the thermal bath to be the same as the FDU temperature. It may also be noted that the eq.(s)(\ref{upth_C1}, \ref{dwnth_C1}) cannot be restored from the eq.(s)(\ref{up_C}, \ref{dwn_C}) even after taking the limit $\alpha d<<1$. Hence, this observation confirms that the equivalence between the transition rates no longer holds for the cases when a two-atom system uniformly accelerates and when a static two-atom system placed in a thermal bath. However, as the ratio of the eqs (\ref{upth_C1}), (\ref{dwnth_C1}), and (\ref{up_C}), (\ref{dwn_C}) matches exactly at $T=\alpha/2\pi$, so the effects of uniform acceleration and the effects of a thermal bath holds completely for the two-atom system interacts with the massless scalar field confined in a cavity.

To obtain the single mirror and free space scenarios, we now take the limiting cases of these expressions. Taking the limit $L\rightarrow\infty$, we find that eq.(s)(\ref{upth_C1}, \ref{dwnth_C1}) reduce to the expression for the upward and the downward transition rate in the presence of a single reflecting boundary
\begin{align}
\mathscr{R}^{\beta}_{\vert \psi\rangle\rightarrow \vert e_{A}e_{B}\rangle}=&\lambda^2\bigg\{\bigg(\frac{\omega_{0}}{2\pi}-\cos^2\theta\,\mathfrak{p}(\omega_{0},\,2z_{0})-\sin^2\theta\,\mathfrak{p}(\omega_{0},\,2(z_{0}+d))\bigg.\bigg.\nonumber\\
&+\left.\bigg.\sin2\theta\Big(\mathfrak{p}\left(\omega_{0},\,d\right)-\mathfrak{p}\left(\omega_{0},\,2z_{0}+d\right)\Big)\bigg)\left(\frac{1}{\exp{(\omega_{0}/T)}-1}\right)\right\}
\end{align}
\begin{align}
\mathscr{R}^{\beta}_{\vert \psi\rangle\rightarrow \vert g_{A}g_{B}\rangle}=&\lambda^2\bigg\{\bigg(\frac{\omega_{0}}{2\pi}-\cos^2\theta\,\mathfrak{p}(\omega_{0},\,2z_{0})-\sin^2\theta\,\mathfrak{p}(\omega_{0},\,2(z_{0}+d))\bigg.\bigg.\nonumber\\
&+\left.\bigg.\sin2\theta\Big(\mathfrak{p}\left(\omega_{0},\,d\right)-\mathfrak{p}\left(\omega_{0},\,2z_{0}+d\right)\Big)\bigg)\left(1+\frac{1}{\exp{(\omega_{0}/T)}-1}\right)\right\}\,.
\end{align}
Similarly, taking the limits $L\rightarrow\infty$ and $z_{0}\rightarrow\infty$, eq.(s)(\ref{upth_C1}, \ref{dwnth_C1}) lead to the expressions for the upward and the downward transition rate in free space given by eq.(s)(\ref{upth_emp}, \ref{dwnth_emp}). 

We now study how the transition rate of an entangled two atom system from an initial entangled state $\vert \psi\rangle$ to a product state with higher energy value $\vert e_{A}e_{B}\rangle$ confined to a cavity varies with parameters such as thermal field temperature ($T$), cavity length ($L$), atomic distance from one boundary ($z_0$), interatomic distance ($d$), and the entanglement parameter ($\theta$). Following the single atom case, here also we present our findings in the following plots. In the following plots, we fix the parameters $T/\omega_0,\,\omega_{0}L,\,\omega_{0}z_{0},\,\text{and}\,\omega_{0}d$ in such a way so that cavity effects remain significant.  

\begin{figure}
\centering
\includegraphics[scale=0.95]{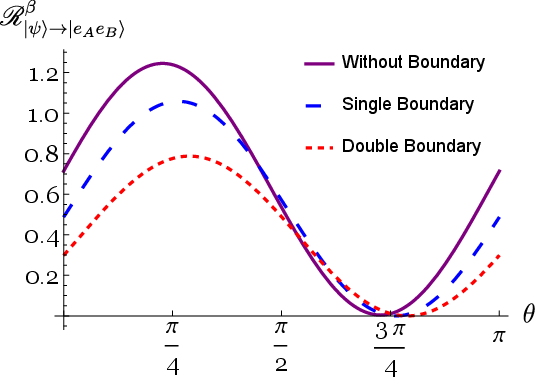}
\caption{Transition rate (per unit $\frac{\lambda^2\omega_0}{2\pi}$) versus entanglement parameter for a fixed value of $\omega_{0}z_0=18,\,\omega_{0}L=20$ (without boundary), $\omega_{0}z_0=1.4,\,\omega_{0}L=8$ (single boundary), and $\omega_{0}z_0=0.6,\,\omega_{0}L=3.4$ (double boundary). All three cases $T/\omega_0=1,\,\omega_{0}d=0.2$.}\label{fig:Double_combth}
\end{figure}

In Figure \ref{fig:Double_combth}, we show the behaviour of the transition rate with respect to the entanglement parameter for the cases where the atoms are in free space, in the vicinity of a single boundary and inside a cavity. From Figure \ref{fig:Double_combth}, it can be seen that the transition rate $\vert \psi\rangle\rightarrow\vert e_{A}e_{B}\rangle$ (per unit $\frac{\lambda^2\omega_0}{2\pi}$) varies sinusoidally with the entanglement parameter $\theta$. In free space, the transition rate increases (from the case corresponding to the zero initial entanglement product state) with increase in the entanglement parameter and it becomes maximum when the initial state is maximally entangled ($\theta=\pi/4$  super-radiant state) \cite{Ficek_2002}. Further increment of the entanglement parameter decreases the transition rate and it vanishes at $\theta=3\pi/4$ (maximally entangled sub-radiant state). In the vicinity of a single boundary, behaviour of the transition rate is quite similar to the free space scenario, with a slight shifting of the extremum points. Inside the cavity, the behaviour of the transition rate is also similar to the free space scenario. The peak value of transition rate inside the cavity is much smaller compared to the free space and single boundary cases. It can be noted that around $\theta=3\pi/4$, the values of the transition rate corresponding to cases of empty space, single boundary, and two boundaries, nearly vanish. At this point, it should be noted that increasing and decreasing the atom-boundary distance $z_0$ and the cavity length $L$ the upward transition rate decreases for the value of  $\theta=\pi/4$.  In the thermal bath, we observe that the transition rate of the superradiant state, $\theta=\pi/4$ of the two atoms inside the cavity does not vanish, in contrast to the result obtained for a co-accelerating frame
in \cite{Mukherjee2023}. Here, the superradiance property of the state ($\theta=\pi/4$) remains intact.  On the other hand, in the thermal bath, we observe that the transition rate of the sub-radiant state, $\theta=3\pi/4$ of the two atoms inside the cavity vanishes. Therefore, entanglement of the initial state is preserved and from a quantum information theoretic viewpoint, this kind of an initial state  may act as a good resource for performing various tasks.

\begin{figure}
\centering
\includegraphics[scale=0.95]{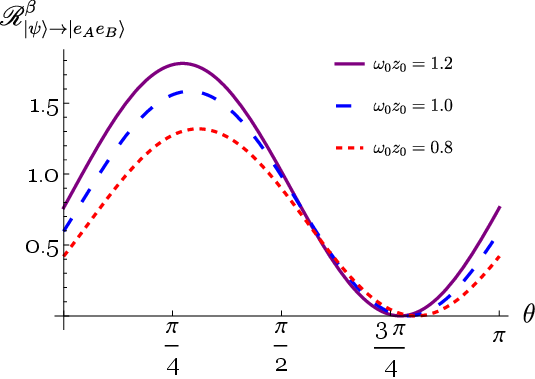}
\caption{Transition rate (per unit $\frac{\lambda^2\omega_0}{2\pi}$) versus entanglement parameter for a fixed value of $T/\omega_0=1,\,\omega_{0}d=0.6,\,\omega_{0}L=3.6$.}\label{fig:Plot_theta1}
\end{figure}

In Figure \ref{fig:Plot_theta1}, we show the behaviour of the transition rate with respect to the entanglement parameter inside a cavity for a fixed value of $T/\omega_0=1,\,\omega_{0}d=0.6,\,\omega_{0}L=3.6$. From Figure \ref{fig:Plot_theta1}, it can be seen that inside the cavity at $\theta=\pi/4$, the upward transition rate $\vert \psi\rangle\rightarrow\vert e_{A}e_{B}\rangle$ (per unit $\frac{\lambda^2\omega_0}{2\pi}$) increases when the distance between one atom and the nearest boundary $z_0$ increases. As the length of the cavity and the inter-atomic distance is fixed therefore after increasing the distance between the nearest atom and the boundary, boundary effect reduces on the number of field modes taking part in the interaction between the atom and the scalar field, which in turn increases the upward transition rates of the system.

\begin{figure}
\centering
\includegraphics[scale=0.95]{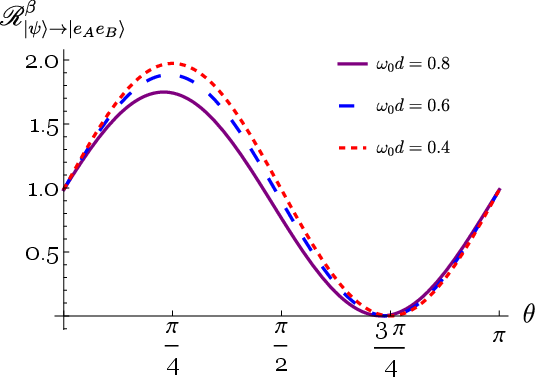}
\caption{Transition rate (per unit $\frac{\lambda^2\omega_0}{2\pi}$) versus entanglement parameter for a fixed value of $T/\omega_0=1,\,\omega_{0}z_0=1.6,\,\omega_{0}L=3.6$.}\label{fig:Plot_theta2}
\end{figure}

In Figure \ref{fig:Plot_theta2}, we show the behaviour of the transition rate with respect to the entanglement parameter inside a cavity for a fixed value of $T/\omega_0=1,\,\omega_{0}z_0=1.6,\,\omega_{0}L=3.6$. From Figure \ref{fig:Plot_theta2}, it can be seen that inside the cavity at $\theta=\pi/4$, the upward transition rate $\vert \psi\rangle\rightarrow\vert e_{A}e_{B}\rangle$ (per unit $\frac{\lambda^2\omega_0}{2\pi}$) decreases when the inter-atomic distance $d$ increases. As the length of the cavity and the distance of one atom from the nearest boundary is fixed therefore after increasing the inter-atomic distance the second atom moves toward the second boundary. Therefore, due to the boundary condition, number of field modes taking part in the inteaction between the atom and the scalar field reduces from the both end and as a result the transition rate of the system decreases. 

\begin{figure}[H]
\centering
\includegraphics[scale=0.95]{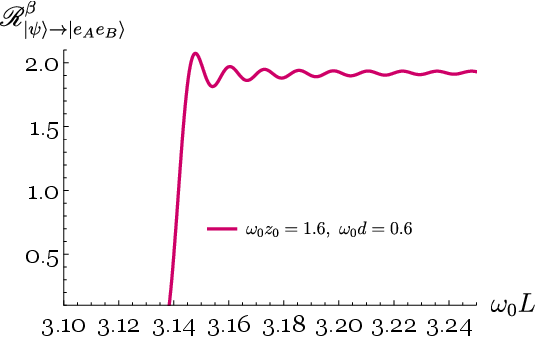}
\caption{Transition rate from $\vert \psi\rangle\rightarrow\vert e_{A}e_{B}\rangle$ (per unit $\frac{\lambda^2\omega_0}{2\pi}$) versus separation between two boundaries for a fixed value of $T/\omega_0=1,\,\omega_0 d=0.5$, and $\omega_0z_0=1.6$.}\label{fig:Double_L}
\end{figure}

\noindent Figure \ref{fig:Double_L} shows the variation of the transition rate from $\vert \psi\rangle\rightarrow\vert e_{A}e_{B}\rangle$ (per unit $\frac{\lambda^2\omega_0}{2\pi}$) with respect to the length of the cavity for a fixed value of the distance of any one atom from one boundary and the interatomic distance. From the figure, it can be seen that for a fixed value of the initial atomic distance $z_0$ of any one atom from the nearest boundary, the transition rate get enhanced when the cavity length increases and attains a maximum value for large values of $L$ ($\omega_0 L>>\omega_0 z_0$). This behaviour is similar
to that of the single atom case, as  mentioned earlier. As more number of field modes take part in the interaction between the scalar field and the  atoms due to the increased cavity length, the transition rate increases. When $\omega_0 L>>\omega_0 z_0$,  the cavity scenario reduces to a single boundary set up and hence the upward transition rate becomes neary constant.

\begin{figure}
\centering
\includegraphics[scale=0.95]{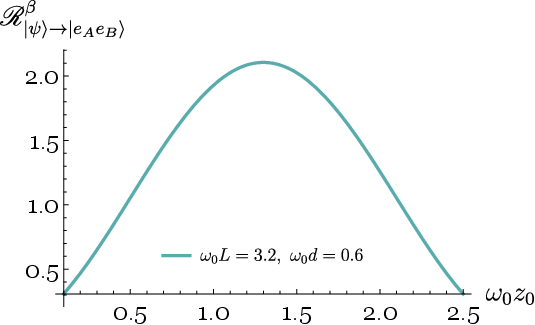}
\caption{Transition rate from $\vert \psi\rangle\rightarrow\vert e_{A}e_{B}\rangle$ (per unit $\frac{\lambda^2\omega_0}{2\pi}$) versus separation between two boundaries for a fixed value of $T/\omega_0=1,\,\omega_0 d=0.5$, and $\omega_0L=3.2$.}\label{fig:Double_z0}
\end{figure}
\noindent Figure \ref{fig:Double_z0} shows the variation of the transition rate from $\vert \psi\rangle\rightarrow\vert e_{A}e_{B}\rangle$ (per unit $\frac{\lambda^2\omega_0}{2\pi}$) with respect to the distance of any one atom from one boundary for a fixed value of the cavity length and the interatomic distance. From the figure, it is observed that for a fixed value of the cavity length $L$ and the interatomic distance $d$, when we increase the atomic distance from one boundary, the transition rate for the two atom system also increases and at a certain value of $z_0$ it attains a maximum value and then it gets reduced by further increment of $z_0$. The reason behind this variation is also similar to the single atom case.

\begin{figure}
\centering
\includegraphics[scale=0.95]{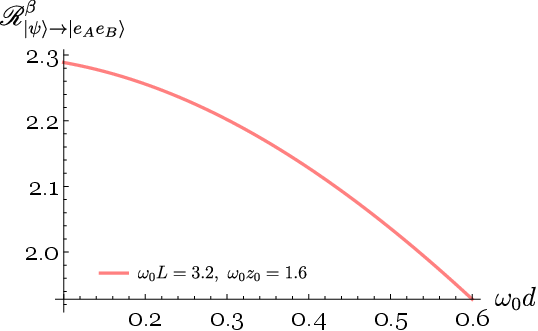}
\caption{Transition rate from $\vert \psi\rangle\rightarrow\vert e_{A}e_{B}\rangle$ (per unit $\frac{\lambda^2\omega_0}{2\pi}$) versus separation between two boundaries, $T/\omega_0=1,\,\omega_0 d=0.5$.}\label{fig:Double_d}
\end{figure}
\noindent Figure \ref{fig:Double_d} shows the variation of the transition rate from $\vert \psi\rangle\rightarrow\vert e_{A}e_{B}\rangle$ (per unit $\frac{\lambda^2\omega_0}{2\pi}$) with respect to the interatomic distance for a fixed value of the cavity length and the distance of any one atom from one boundary. From the figure, it is observed that for a fixed value of the cavity length $L$ and the distance of any one atom from one boundary $z_0$, when we increase the interatomic distance, the transition rate for the two atom system decreases. Due to the increament of the interatomic distance the second atom moves toward the second boundary. Therefore, due the boundary effect the transition rate of the two atom system falls down. 

\begin{figure}[H]
\centering
\includegraphics[scale=0.9]{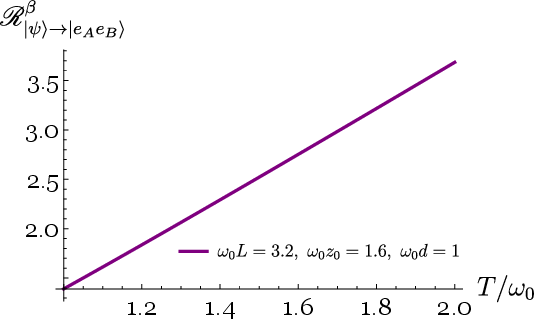}
\caption{Transition rate from $\vert \psi\rangle\rightarrow\vert e_{A}e_{B}\rangle$ (per unit $\frac{\lambda^2\omega_0}{2\pi}$) versus acceleration, $\omega_{0}d=0.5,\,\omega_{0}z_0=0.4$.}\label{fig:Double_T}
\end{figure}

\noindent Figure \ref{fig:Double_T} shows the variation of the transition rate from $\vert \psi\rangle\rightarrow\vert e_{A}e_{B}\rangle$ (per unit $\frac{\lambda^2\omega_0}{2\pi}$) with respect to the atomic acceleration for a fixed value of the cavity length. From the figure, it is observed that for a fixed value of the length of the cavity $L$, the atomic distance $z_0$ from one boundary, and the interatomic distance $d$, it is observed that the transition rate increases when the temperature of the thermal bath is increased. 
\section{Quantitative estimation of the transition rate}\label{sec:Ex}
In this section, we now present a quantitative estimation of the upward transition rate of a single and two atom system inside a cavity by considering the Rubidium and Caesium atom. Following some recent works Ref. \cite{corzo2019waveguide, Vetsch2010, Goban2012, Vylegzhanin_2023, Donaire, 
Urban2009}, it has been seen that $^{87}Rb$ and $^{133}Cs$ widely used in the experimental setup. 
\subsection{Transition rate of $^{87}Rb$}
From the Ref. \cite{Steck2024Rubidium, Vylegzhanin_2023}, we have taken the transition energy and the cavity length for calculating the upward transition rate of $^{87}Rb$. The values are $\hbar\omega_0=1.59$eV and $L=400$nm respectively. Setting thermal bath temperature $T=20,000$K and the distance between the atom and the nearest boundary $z_0=100$nm. Fixing all the physical quantity and calculating the dimensionless parameters, we get
\begin{equation}
T/\omega_0=1,\hspace{2cm} \omega_0L=3.2,\hspace{2cm}\omega_0z_0=0.8
\end{equation}
Taking the coupling constant $\lambda=0.1$ and using above values of the parameters, the upward transition rate for a single $^{87}Rb$ atom becomes
\begin{equation}
\mathscr{R}^{\beta}_{\vert g\rangle\rightarrow \vert e\rangle}=1.45\times 10^{-3}\,\text{eV}=3.51\times 10^{11}\,\text{s}^{-1}.
\end{equation}
In case of two $^{87}Rb$ atoms, we additionally choose the inter-atomic distance $d=50$nm and the entanglement parameter $\theta=\pi/4$. Therefore, $\omega_0d=0.4$. Taking $\lambda=0.1$, the upward transition rate for two $^{87}Rb$ atoms becomes
\begin{equation}
\mathscr{R}^{\beta}_{\vert \psi\rangle\rightarrow \vert e_{A}e_{B}\rangle}=3.87\times 10^{-3}\,\text{eV}=9.37\times 10^{11}\,\text{s}^{-1}.
\end{equation}
\subsection{Transition rate of $^{133}Cs$}
From the Ref. \cite{Steck2024Cesium, Vylegzhanin_2023}, we have taken the transition energy and the cavity length for calculating the upward transition rate of $^{133}Cs$. The values are $\hbar\omega_0=1.46$eV and $L=500$nm respectively. Setting thermal bath temperature $T=20,000$K and the distance between the atom and the nearest boundary $z_0=150$nm. Fixing all the physical quantity and calculating the dimensionless parameters, we get
\begin{equation}
T/\omega_0=1.2,\hspace{2cm} \omega_0L=3.7,\hspace{2cm}\omega_0z_0=1.1
\end{equation}
Taking the coupling constant $\lambda=0.1$ and using above values of the parameters, the upward transition rate for a single $^{133} Cs$ atom becomes
\begin{equation}
\mathscr{R}^{\beta}_{\vert g\rangle\rightarrow \vert e\rangle}=1.96\times 10^{-3}\,\text{eV}=4.74\times 10^{11}\,\text{s}^{-1}.
\end{equation}
In case of two $^{133}Cs$ atoms, we additionally choose the inter-atomic distance $d=75$nm and the entanglement parameter $\theta=\pi/4$. Therefore, $\omega_0d=0.4$. Taking $\lambda=0.1$, the upward transition rate for two $^{133}Cs$ atoms becomes
\begin{equation}
\mathscr{R}^{\beta}_{\vert \psi\rangle\rightarrow \vert e_{A}e_{B}\rangle}=4.86\times 10^{-3}\,\text{eV}=1.18\times 10^{12}\,\text{s}^{-1}.
\end{equation}
\section{Conclusions}\label{sec:Con}
 In this work we take a fresh look at the FDU effect in the context of two-level single and entangled atomic systems that are static in a thermal bath. We investigated the question as to whether the FDU effect shows up or not for single atom and entangled two-atom detector systems in free-space and inside cavity setups inside a thermal bath. In order to investigate the actual equivalence in real physical systems, the first step is to identify choices of specific detectors in context of which one may hope to observe  such equivalence. Phenomena such as the excitation rate of a detector are directly observable, and are therefore of substantial significance in empirically verifying deep physical concepts, such as the FDU effect.
 

Our study takes into consideration both single and entangled two-atom systems which can be located either in free space or inside cavities with reflecting boundary conditions. The two-atom system is taken to be initially prepared in a most general pure entangled state. The interactions of both systems with a massless scalar field in its vacuum state are studied in order to investigate the equivalence between the thermal bath and the uniform acceleration for the cases of the single and two-atom systems. For doing so, we have calculated the upward and downward transition rates of the single and entangled atomic systems immersed in a thermal bath in free space and inside cavities. We have compared these results with those of the results of accelerated single and entangled atomic systems in free space and inside a cavity \cite{Mukherjee2023} given in Appendix \ref{Appendix:AA}.


From the results, it is observed that while both the upward and downward transitions occur for the single atom as well as the entangled two-atom system, the upward transition is significant as it solely arises due to the temperature of the thermal bath. The actual transition rate depends on the cavity  length, the distance of an atom from one boundary, the temperature of the thermal bath, and the magnitude of initial atomic entanglement.  From our analysis it is evident that the transition rate shows oscillatory behaviour with the entanglement parameter. Considering a small magnitude of initial entanglement, we find that  increasing the entanglement parameter enhances the upward transition and downward transition rates in free space, whereas, both the transition rates get suppressed due to the decreased number of field modes participating in the interaction between the atom and the scalar field in the presence of the cavity. In the case when the initial entanglement parameter has the value $\theta=3\pi/4$, we observe that both the transition rates vanish, indicating that no transition occurs from the maximally entangled initial state to any higher or lower energetic product state. Hence, the entanglement of the sub-radiant initial state can be preserved, a result which may be of significance, since preservation of entanglement enables its use as a resource for performing various quantum information processing tasks.


Our  investigation of the upward and downward transition rates enables us to conclude that in the case of single as well as two-atom systems, the physical state of the system and the observer's reference frame both significantly influence the upward and downward transition rates of the systems. 
Whether the equivalence inherent in the FDU effect is manifested or not depends upon the interplay of the following physical conditions. 
In free space, the upward and downward transition rates of a static atom placed in a thermal bath match with those of an accelerated atom, if the temperature of the thermal bath is equal to the Unruh temperature $T=\alpha/2\pi$. Such kind of equivalence also holds at the level of transition rates when a single atom is placed in a coaccelerated frame in a thermal bath \cite{Mukherjee2023}. The picture for two atoms is more intricate where atomic entanglement plays the key role. In an earlier work, it is observed that the upward and downward transition rates of two entangled atoms accelerating through a cavity turns out to be similar to the transition rates of a coaccelerated observer in a thermal bath \cite{Mukherjee2023}. However, in this study it is seen that the upward and downward transition rates of a uniformly accelerated two-atom system with the transition rates when the system is static but immersed in a thermal bath, such equivalence between the transition rates holds only under specific limiting conditions in free space, but breaks down completely inside a cavity setup. However, it turns out that the ratio of the upward and downward transition rates inside the thermal bath in free space as well as inside a cavity match exactly with those of the accelerated system in free space and inside a cavity.

Apart from the computing of the transition rate in different scenario for single and two level system, in this study we provide a quantitative estimation of the transition rate inside the cavity setup. Following some recent experimental setup \cite{corzo2019waveguide, Vetsch2010, Goban2012, Vylegzhanin_2023, Donaire, 
Urban2009}, we fix the physical parameters for the single and two Rubidium ($^{87}Rb$) and Cesium ($^{133}Cs$) atoms respectively. From this estimation, it can be seen that within the cavity of order nanometer the transition rates of this atoms are very high.


Our present analysis serves to comprehensively reestablish that in general, there is no equivalence at the level of transition rates between the accelerating and the static thermal bath frame for the chosen single atom model inside a cavity and the entangled two-atom detector model as realized in both free-space and cavity setups. However, at the level of the ratio of the transition rates, there is a complete equivalence between the accelerating and the static thermal bath frame for both the single and the entangled two-atom detector model as realized in both free-space and cavity setups. Therefore, at this point we should emphasize though, that our present analysis does not impact in any way the conceptual validity of the FDU effect. Rather, its revelation through the manifestation of equivalence, is model dependent \cite{Crispino2}, as evident through comparison of earlier results involving co-accelerating observers \cite{Zhou2020, Mukherjee2023}.


Finally, it may be noted that in our present study we have considered only the effect of varying magnitudes of initial atomic entanglement on the transition rates. The focus here is not to evaluate any effects  of generation or degradation of entanglement that may arise from the dynamics. However, other phenomena such as generation or degradation of atomic entanglement, as well as atom-field entanglement may have interesting implications \cite{ASM_2001, ASM_2006, Gupta_2018, Jiawei_2015, Soares_2022}. Evaluation of such effects could be undertaken through  the master-equation approach \cite{GKS_1976, Lindblad_1976, Benatti_2004} in future works.
\section*{Acknowledgement}
AM and ASM acknowledges support from project no. DST/ICPS/QuEST/2019/Q79 of the Department of Science and Technology (DST), Government of India.
\begin{appendices}
\section{Interaction of the accelerated atomic system with a massless scalar field}\label{Appendix:AA}
In this Appendix, we mainly review some key results of \cite{Mukherjee2023} which we use for the sake of comparison with the 
subsequent results of our present work.

\subsection{Single atom system}\label{sec:CoupS}

Let us consider a single atom (an Unruh-DeWitt detector) with two energy levels $\vert g\rangle$ and $\vert e\rangle$ with corresponding energy values $-\omega_{0}/2$ and $+\omega_{0}/2$, travelling along a stationary trajectory in a vacuum with massless scalar field fluctuations. In the laboratory frame, trajectories of the atom can be represented through $x(\tau)\equiv (t(\tau),\mathbf{x}(\tau))$. In the instantaneous inertial frame, the Hamiltonian describing the atom-field interaction in the interaction picture is given by \cite{Svaiter1}
\begin{equation}
\mathscr{H}_{int} = \lambda m(\tau) \phi(x(\tau))\,\label{hamS}
\end{equation}
where $\lambda$ is a small coupling constant, $m(\tau)=e^{i \mathscr{H}_{0}\tau}m(0)e^{-i \mathscr{H}_{0}\tau}$ is the monopole operator at any proper time $\tau$ of a single atom, $\phi(x(\tau))$ is the massless quantum scalar field evaluated at the trajectory $x(\tau)$ with
$m(0)=\vert g\rangle\langle e\vert +\vert e\rangle\langle g\vert$
being the initial monopole operator and $\mathscr{H}_0\vert e\rangle=\frac{\omega}{2}\vert e\rangle$ being the free Hamiltonian of a single atom respectively \cite{Mukherjee2022}.\\
Using the formalism discussed in \cite{birrell1984quantum, Mukherjee2023}, the rate of transition probability from the initial atomic state $\vert i\rangle$ to the final atomic state $\vert f\rangle$ turns out to be
\begin{equation}
\mathscr{R}_{\vert i\rangle\rightarrow \vert f \rangle}=\lambda^2 \vert m_{fi}\vert^2 \mathscr{F}(\Delta E)\,\label{TransprobrateS}
\end{equation}
where $\Delta E= E_{f}-E_{i}$, $m_{fi}=\langle f\vert m(0)\vert i\rangle$, and the response function per unit proper time can be written as
\begin{equation}
\mathscr{F}(\Delta E)=\int_{-\infty}^{+\infty} d(\Delta\tau) \,e^{-i\Delta E\Delta\tau}\,G^{+}(x(\tau),x(\tau'))\label{ResponserateS}
\end{equation}
where $\Delta\tau=\tau-\tau'$ and
\begin{equation}
G^{+}(x(\tau),x(\tau'))=\langle 0_{M}\vert\phi(x(\tau))\phi(x(\tau'))\vert 0_{M}\rangle\label{WightmanS}
\end{equation}
is the positive frequency Wightman function of the massless scalar field \cite{birrell1984quantum}.\\
In empty space, using the atomic trajectory in the laboratory frame $t(\tau)=\frac{1}{\alpha}\sinh(\alpha\tau)$, $x(\tau)=\frac{1}{\alpha}\cosh(\alpha\tau)$, $y=z=0$
with $\alpha$ being the proper acceleration and $\tau$ being the proper time of the atom, the Wightman function becomes
\begin{equation}
 G^{+}(x(\tau),x(\tau'))=-\frac{\alpha^2}{16\pi^2}\frac{1}{\sinh^2\left[\frac{1}{2}(\alpha\Delta\tau-i\varepsilon)\right]}\,. \label{Wght1S}
\end{equation}
Substituting the above Wightman function into eq.\eqref{TransprobrateS}, the upward and downward transition rate becomes (See \cite{Mukherjee2023} for a detailed calculation)
\begin{equation}\label{Up_empS}
\mathscr{R}_{\vert g\rangle\rightarrow \vert e\rangle}=\frac{\lambda^2\omega_0}{2\pi}\left(\frac{1}{\exp(2\pi\omega_0/\alpha)-1}\right)
\end{equation}
\begin{equation}\label{Down_empS}
\mathscr{R}_{\vert e\rangle\rightarrow \vert g\rangle}=\frac{\lambda^2\omega_0}{2\pi}\left(1+\frac{1}{\exp(2\pi\omega_0/\alpha)-1}\right)\,.
\end{equation}
From the above equations, it is clearly seen that in case of a single atom system the upward transition rate solely depends on the atomic acceleration. Taking the ratio of the above two results, we get
\begin{equation}\label{Ratio0}
\frac{\mathscr{R}_{\vert g\rangle\rightarrow \vert e\rangle}}{\mathscr{R}_{\vert e\rangle\rightarrow \vert g\rangle}}\equiv\frac{\mathscr{R}_{up}}{\mathscr{R}_{down}}=\exp(-2\pi\omega_0/\alpha)\,.
\end{equation}

\begin{figure}[H]
\centering
\includegraphics[scale=0.4]{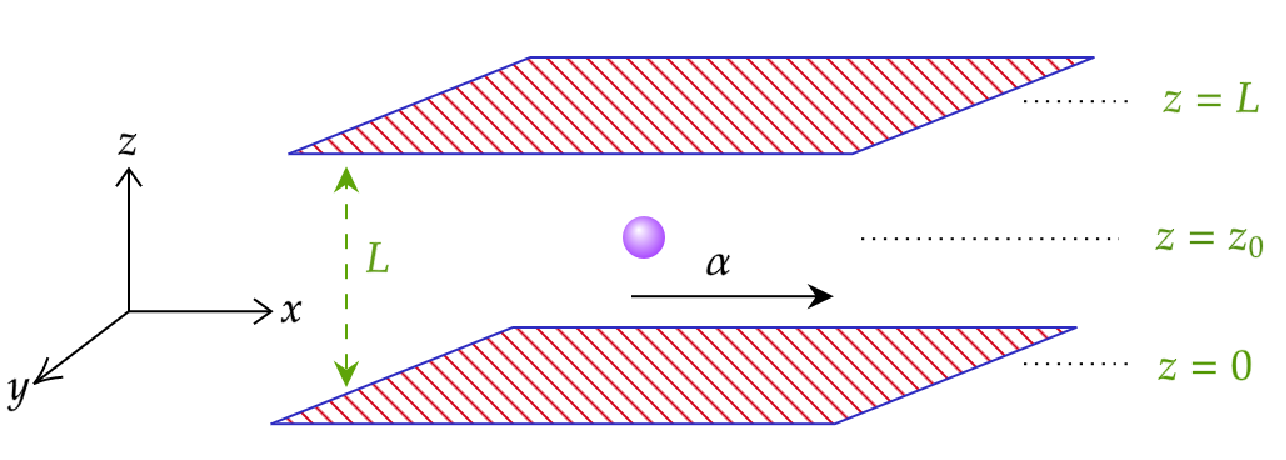}
\caption{Uniformly accelerated atom confined in a cavity \cite{Mukherjee2023}.}\label{fig:SingleC_new}
\end{figure}

\noindent Now, inside a cavity of length $L$ as shown in Figure \ref{fig:SingleC_new}, with the atomic trajectory being $t(\tau)=\frac{1}{\alpha}\sinh(\alpha\tau)$, $x(\tau)=\frac{1}{\alpha}\cosh(\alpha\tau)$, $y=0$, and $z=z_{0}$, the Wightman function is given by \cite{Mukherjee2023}
\begin{equation}
 G^{+}(x(\tau),x(\tau'))=-\frac{\alpha^2}{16\pi^2}\displaystyle\sum_{n=-\infty}^{\infty}\left[\frac{1}{\sinh^2\left[\frac{1}{2}(\alpha\Delta\tau-i\varepsilon)\right]-\frac{1}{4}d_{1}^2\alpha^2}-\frac{1}{\sinh^2\left[\frac{1}{2}(\alpha\Delta\tau-i\varepsilon)\right]-\frac{1}{4}d_{2}^2\alpha^2}\right] \label{WghtC1S}
\end{equation}
with $d_1=nL,\,d_2=2z_{0}-nL$.
Now using the process outlined in Appendix B of \cite{Mukherjee2023}, the upward and downward transition rate inside the cavity turn out to be
\begin{equation}\label{Up_CS}
\mathscr{R}_{\vert g\rangle\rightarrow \vert e\rangle}=\lambda^2\left[\left\{\frac{\omega_0}{2\pi}+\mathfrak{f}\left(\omega_0,\,\alpha,\,\frac{L}{2}\right)-\mathfrak{h}\left(\omega_0,\,\alpha,\,z_{0},\frac{L}{2}\right)\right\}\left(\frac{1}{\exp{(2\pi\omega_0/\alpha)}-1}\right)\right]
\end{equation}
\begin{equation}\label{Down_CS}
\mathscr{R}_{\vert e\rangle\rightarrow \vert g\rangle}=\lambda^2\left[\left\{\frac{\omega_0}{2\pi}+\mathfrak{f}\left(\omega_0,\,\alpha,\,\frac{L}{2}\right)-\mathfrak{h}\left(\omega_0,\,\alpha,\,z_{0},\frac{L}{2}\right)\right\}\left(1+\frac{1}{\exp{(2\pi\omega_0/\alpha)}-1}\right)\right]\,
\end{equation}
where we have defined
\begin{align}
\mathfrak{f}\left(\Delta E,\,\alpha,\,\frac{L}{2}\right)&=2\displaystyle\sum_{n=1}^{\infty}\mathfrak{g}\left(\Delta E,\,\alpha,\,\frac{nL}{2}\right)\label{ffunc1S}\\
\mathfrak{h}\left(\Delta E,\,\alpha,\,z_0\,,\frac{L}{2}\right)&=\displaystyle\sum_{n=-\infty}^{\infty}\mathfrak{g}\left(\Delta E,\,\alpha,\,z_0-\frac{nL}{2}\right)\,\label{hfunc1S}
\end{align}
with $\mathfrak{g}\left(\Delta E,\,\alpha,\,z_0\right)$  defined as
\begin{equation}
\mathfrak{g}\left(\Delta E,\,\alpha,\,z_0\right)=\frac{\sin(\frac{2\Delta E}{\alpha}\sinh^{-1}(\alpha z_0))}{4\pi z_0\sqrt{1+\alpha^2 z_0^2}}\,.\label{gfuncS}
\end{equation}
Taking the ratio of the eqs.(\ref{Up_CS}) and (\ref{Down_CS}), in cavity scenario, we also get
\begin{equation}\label{Ratio1}
\frac{\mathscr{R}_{\vert g\rangle\rightarrow \vert e\rangle}}{\mathscr{R}_{\vert e\rangle\rightarrow \vert g\rangle}}\equiv\frac{\mathscr{R}_{up}}{\mathscr{R}_{down}}=\exp(-2\pi\omega_0/\alpha)\,.
\end{equation}
\subsection{Two-atom system}\label{sec:CoupT}
In this subsection, considering two identical atoms $A$ and $B$, we assume that they are travelling synchronously along stationary trajectories in the vacuum of a  massless scalar field. The interatomic distance is assumed to be constant and the proper times of the two atoms can be described by the same time $\tau$ \cite{Zhang2019}. In the laboratory frame, trajectories of the two atoms can be represented through $x_{A}(\tau)$ and $x_{B}(\tau)$. Here we consider each atom as a two level system having energy levels $\vert g\rangle$ and $\vert e\rangle$ with corresponding energy values $-\omega_{0}/2$ and $+\omega_{0}/2$. Therefore, the entire two-atom system can be described by the eigenstates of the initial {\it atomic Hamiltonian}, namely, $\vert g_{A},g_{B}\rangle$, $\vert g_{A},e_{B}\rangle$, $\vert e_{A},g_{B}\rangle$, and $\vert e_{A},e_{B}\rangle$ with the corresponding energy eigenvalues $-\omega_{0},\,0,\,0,\,\,\text{and}\,\,\omega_{0}$ respectively. Here, as the eigenstates $\vert g_{A},e_{B}\rangle$ and $\vert e_{A},g_{B}\rangle$ are the degenerate eigenstates of the Hamiltonian with the energy value zero, therefore any linear combination of this eigenstates will also be an eigenstate of this system with same energy eigenvalue \cite{Svaiter2}. Hence, the most general quantum state of the two-atom system with zero energy value is given by \cite{Chatterjee2021_1}
\begin{equation}\label{state_psi}
\vert \psi\rangle=\sin\theta\vert g_{A},e_{B}\rangle + \cos\theta \vert e_{A},g_{B}\rangle
\end{equation}
where the entanglement parameter $\theta$ lies in the range $0\leq\theta\leq\pi$. It may be noted  that though the Bell-state basis is frequently used when working with entangled states, for our present purpose use of the Bell basis is inconvenient,  since the Bell states are not eigenstates of the atomic Hamiltonian.

In the instantaneously inertial frame, the Hamiltonian describing the atom-field interaction is given by
\begin{equation}
\mathscr{H}_{int} = \lambda \displaystyle\sum_{\xi=A,\,B} m_{\xi}(\tau) \phi(x_{\xi}(\tau))\,\label{ham}
\end{equation}
where $\lambda$ is a small atom-field coupling constant.

As a result of the atom-field interaction, the transition probability rate of the two-atom system from the initial state $\vert \chi\rangle$ to the final state $\vert \chi'\rangle$ turns out to be
\begin{equation}
\mathscr{R}_{\vert \chi\rangle\rightarrow \vert \chi' \rangle}=\lambda^2 \Big[\vert m^{(A)}_{\chi'\chi}\vert^2 \mathscr{F}_{AA}(\Delta E)+m^{(B)}_{\chi'\chi}\,m^{(A)\,*}_{\chi'\chi}\mathscr{F}_{AB}(\Delta E)\Big]+A\rightleftharpoons B\,\text{terms}\,\label{transprobrate}
\end{equation}
where $m^{(A)}_{\chi'\chi}=\langle\chi'\vert m(0)\otimes \mathds{1}_{B}\vert\chi\rangle$,  $m^{(B)}_{\chi'\chi}=\langle\chi'\vert \mathds{1}_{A}\otimes m(0)\vert\chi\rangle$.
The response function per unit proper time can be written as
\begin{equation}
\mathscr{F}_{\xi\xi'}(\Delta E)=\int_{-\infty}^{+\infty} d(\Delta\tau) \,e^{-i\Delta E\Delta\tau}\,G^{+}(x_{\xi}(\tau),x_{\xi'}(\tau'))\label{responserate}
\end{equation}
where $\Delta\tau=\tau-\tau'$, $\xi,\xi'$ can be labeled by $A$ or $B$, and 
\begin{equation}
G^{+}(x_{\xi}(\tau),x_{\xi'}(\tau'))=\langle 0_{M}\vert\phi(x_{\xi}(\tau))\phi(x_{\xi'}(\tau'))\vert 0_{M}\rangle\label{wightman}
\end{equation}
is the Wightman function of the massless scalar field.

In empty space, using the trajectories of both the atoms in the laboratory frame $t_{A}(\tau)=t_{B}(\tau)=\frac{1}{\alpha}\sinh(\alpha\tau)$, $x_{A}(\tau)=x_{B}(\tau)=\frac{1}{\alpha}\cosh(\alpha\tau)$, $y_{B}=y_{A}+d$, and $z_{A}=z_{B}=0$,
with $d$ being the constant interatomic distance, $\alpha$ being the proper acceleration and $\tau$ being the proper time of the two-atom system, the transition rates of the two-atom system
from the initial entangled state $\vert \psi\rangle$ to the final product states $\vert e_{A},\,e_{B}\rangle$ and $\vert g_{A},\,g_{B}\rangle$ can be expressed as \cite{Mukherjee2023}
\begin{equation}\label{up_emp}
\mathscr{R}_{\vert \psi\rangle\rightarrow \vert e_{A},\,e_{B}\rangle}=\lambda^2\left\{\left(\frac{\omega_{0}}{2\pi}+\frac{\sin 2\theta \sin(\frac{2\omega_{0}}{\alpha}\sinh^{-1}(\frac{1}{2}\alpha d))}{2\pi d\sqrt{1+\frac{1}{4}d^2\alpha^2}}\right)\left(\frac{1}{\exp{(2\pi\omega_{0}/\alpha)}-1}\right)\right\}\,
\end{equation}

\begin{equation}\label{dwn_emp}
\mathscr{R}_{\vert \psi\rangle\rightarrow \vert g_{A},\,g_{B}\rangle}=\lambda^2\left\{\left(\frac{\omega_{0}}{2\pi}+\frac{\sin 2\theta \sin(\frac{2\omega_{0}}{\alpha}\sinh^{-1}(\frac{1}{2}\alpha d))}{2\pi d\sqrt{1+\frac{1}{4}d^2\alpha^2}}\right)\left(1+\frac{1}{\exp{(2\pi\omega_{0}/\alpha)}-1}\right)\right\}\,.
\end{equation}
In the limit $\alpha d<<1$, above equations upto $\mathcal{O}(\alpha^2 d^2)$ take the form
\begin{align}\label{up_empA}
\mathscr{R}_{\vert \psi\rangle\rightarrow \vert e_{A},\,e_{B}\rangle}= &\lambda^2\left\{\left(\frac{\omega_{0}}{2\pi}+\frac{\sin 2\theta}{2\pi d}\left[\sin(\omega_0 d)-\frac{1}{8}\left\{\sin(\omega_0 d)+\frac{1}{3}(\omega_0 d)\cos(\omega_0 d)\right\}\alpha^2 d^2\right]\right)\right.\nonumber\\
&\times\left.\left(\frac{1}{\exp{(2\pi\omega_{0}/\alpha)}-1}\right)\right\}\,
\end{align}

\begin{align}\label{dwn_empA}
\mathscr{R}_{\vert \psi\rangle\rightarrow \vert g_{A},\,g_{B}\rangle}= &\lambda^2\left\{\left(\frac{\omega_{0}}{2\pi}+\frac{\sin 2\theta}{2\pi d}\left[\sin(\omega_0 d)-\frac{1}{8}\left\{\sin(\omega_0 d)+\frac{1}{3}(\omega_0 d)\cos(\omega_0 d)\right\}\alpha^2 d^2\right]\right)\right.\nonumber\\
&\times\left.\left(1+\frac{1}{\exp{(2\pi\omega_{0}/\alpha)}-1}\right)\right\}\,.
\end{align}
\begin{figure}[!htbp]
\centering
\includegraphics[scale=0.4]{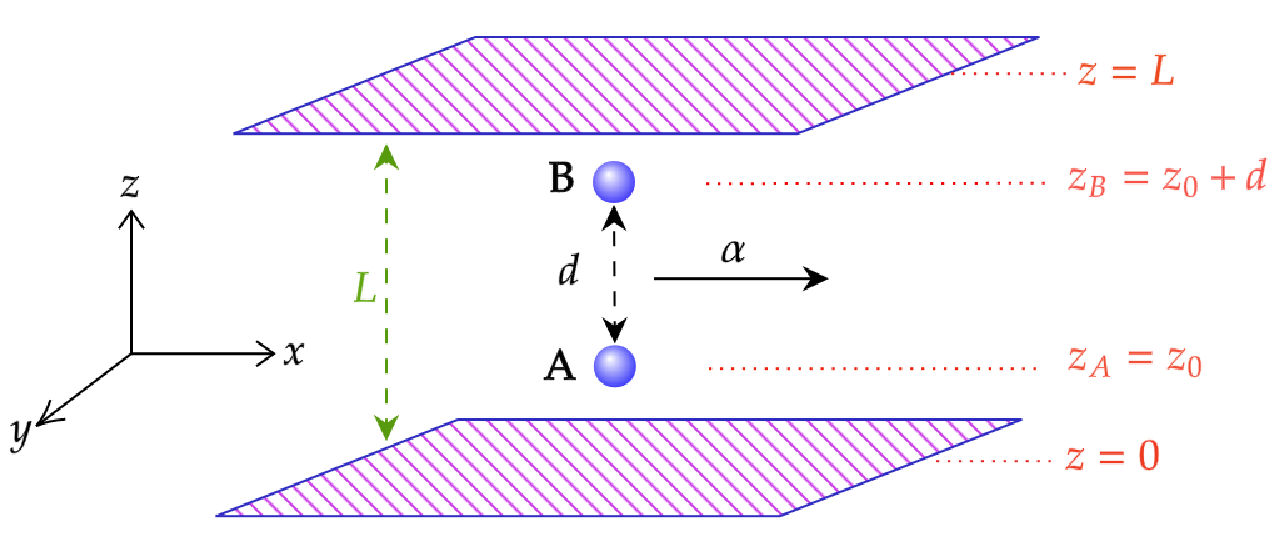}
\caption{Uniformly accelerated two-atom confined in a cavity \cite{Mukherjee2023}.}\label{fig:DoubleC_new}
\end{figure}
\noindent Inside a cavity of length $L$ as shown in Figure \ref{fig:DoubleC_new}, using the atomic trajectories $t_{A/B}(\tau)=\frac{1}{\alpha}\sinh(\alpha\tau)$, $x_{A/B}(\tau)=\frac{1}{\alpha}\cosh(\alpha\tau)$, $y_{A/B}=y_{0}$, and $z_{A}=z_{0},\,\,z_{B}=z_{0}+d$, the transition rate of the two-atom system
from the initial entangled state $\vert \psi\rangle$ to the final product state $\vert e_{A},\,e_{B}\rangle$ and $\vert g_{A},\,g_{B}\rangle$ inside the cavity can be expressed as \cite{Mukherjee2023}
\begin{align}\label{up_C}
\mathscr{R}_{\vert \psi\rangle\rightarrow \vert e_{A},\,e_{B}\rangle}=&\lambda^2\left\{\left(\frac{\omega_0}{2\pi}+\mathfrak{f}\left(\omega_{0},\,\alpha,\,\frac{L}{2}\right)-\cos^2\theta\,\mathfrak{h}\left(\omega_{0},\,\alpha,\,z_{0},\frac{L}{2}\right)-\sin^2\theta\,\mathfrak{m}\left(\omega_{0},\,\alpha,\,z_{0},\,d,\,\frac{L}{2}\right)\right.\right.\nonumber\\
&+\left.\sin2\theta\,\left(\mathfrak{n}\left(\omega_{0},\,\alpha,\,\frac{d}{2},\,\frac{L}{2}\right)-\mathfrak{m}\left(\omega_{0},\,\alpha,\,z_{0},\,\frac{d}{2},\,\frac{L}{2}\right)\right)\left(\frac{1}{\exp{(2\pi\omega_{0}/\alpha)}-1}\right)\right\}
\end{align}

\begin{align}\label{dwn_C}
\mathscr{R}_{\vert \psi\rangle\rightarrow \vert g_{A},\,g_{B}\rangle}=&\lambda^2\left\{\left(\frac{\omega_0}{2\pi}+\mathfrak{f}\left(\omega_{0},\,\alpha,\,\frac{L}{2}\right)-\cos^2\theta\,\mathfrak{h}\left(\omega_{0},\,\alpha,\,z_{0},\frac{L}{2}\right)-\sin^2\theta\,\mathfrak{m}\left(\omega_{0},\,\alpha,\,z_{0},\,d,\,\frac{L}{2}\right)\right.\right.\nonumber\\
&+\left.\sin2\theta\,\left(\mathfrak{n}\left(\omega_{0},\,\alpha,\,\frac{d}{2},\,\frac{L}{2}\right)-\mathfrak{m}\left(\omega_{0},\,\alpha,\,z_{0},\,\frac{d}{2},\,\frac{L}{2}\right)\right)\left(1+\frac{1}{\exp{(2\pi\omega_{0}/\alpha)}-1}\right)\right\}
\end{align}
where we have defined
\begin{align}
\mathfrak{m}\left(\Delta E,\,\alpha,\,z_0\,,d,\,\frac{L}{2}\right)&=\displaystyle\sum_{n=-\infty}^{\infty}\mathfrak{g}\left(\Delta E,\,\alpha,\,z_0+d-\frac{nL}{2}\right)\,\label{mfunc}\\
\mathfrak{n}\left(\Delta E,\,\alpha,\,\frac{d}{2},\,\frac{L}{2}\right)&=\displaystyle\sum_{n=-\infty}^{\infty}\mathfrak{g}\left(\Delta E,\,\alpha,\,\frac{d-nL}{2}\right)\,\label{nfunc}
\end{align}
with $\mathfrak{f}\left(\Delta E,\,\alpha,\,\frac{L}{2}\right),\,\mathfrak{h}\left(\Delta E,\,\alpha,\,z_0,\,\frac{L}{2}\right),\,\mathfrak{g}\left(\Delta E,\,\alpha,\,z_0\right)$ are given in eq.(s)(\ref{ffunc1S}, \ref{hfunc1S}, \ref{gfuncS}).
\section{Derivation of the thermal Wightman function}\label{Appendix:BB}
In this Appendix, for the sake of completeness, we provide a complete derivation of the thermal Wightman function of the massless scalar field eq.\eqref{wightmn1S}.\\
Thermal Wightman function is defined as
\begin{equation}
G^{+}_{\beta}(x(\tau),x(\tau'))=\frac{tr[e^{-\beta \mathscr{H}_{F}}\phi(x(\tau))\phi(x(\tau'))]}{tr[e^{-\beta \mathscr{H}_{F}}]}\label{A1}
\end{equation}
where $\mathscr{H}_F=\displaystyle\sum_{\mathbf{k}}\omega_\mathbf{k} a^{\dagger}_\mathbf{k} a_\mathbf{k}$.\\
Therefore,
\begin{align}
&G^{+}_{\beta}(x(\tau),x(\tau'))=\frac{tr[e^{-\beta \mathscr{H}_{F}}\phi(x(\tau))\phi(x(\tau'))]}{tr[e^{-\beta \mathscr{H}_{F}}]}\nonumber\\
=&tr\left[\exp\left\{-\beta \displaystyle\sum_{\mathbf{k}}\omega_\mathbf{k} a^{\dagger}_\mathbf{k} a_\mathbf{k}\right\}\phi(x(\tau))\phi(x(\tau'))\right]\bigg/tr\left[\exp\left\{-\beta \displaystyle\sum_{\mathbf{k}}\omega_\mathbf{k} a^{\dagger}_\mathbf{k} a_\mathbf{k}\right\}\right]\nonumber\\
=&\displaystyle\sum_{\xi=0}^{\infty}\langle\xi\vert\exp\left\{-\beta \displaystyle\sum_{\mathbf{k}}\omega_\mathbf{k} a^{\dagger}_\mathbf{k} a_\mathbf{k}\right\}\phi(x(\tau))\phi(x(\tau'))\vert \xi\rangle\bigg/\displaystyle\sum_{\xi=0}^{\infty}\langle\xi\vert\exp\left\{-\beta \displaystyle\sum_{\mathbf{k}}\omega_\mathbf{k} a^{\dagger}_\mathbf{k} a_\mathbf{k}\right\}\vert\xi\rangle\nonumber\\
=&\displaystyle\sum_{\xi,\sigma=0}^{\infty}\langle\xi\vert\exp\left\{-\beta \displaystyle\sum_{\mathbf{k}}\omega_\mathbf{k} a^{\dagger}_\mathbf{k} a_\mathbf{k}\right\}\vert\sigma\rangle\langle\sigma\vert\phi(x(\tau))\phi(x(\tau'))\vert \xi\rangle\bigg/\displaystyle\sum_{\xi=0}^{\infty}e^{-\beta\xi\omega}\nonumber\\
=&\bigg[\displaystyle\sum_{\sigma=0}^{\infty}exp(-\beta \sigma\omega)\langle\sigma\vert\phi(x(\tau))\phi(x(\tau'))\vert \sigma\rangle\bigg]\bigg/\displaystyle\sum_{\sigma=0}^{\infty}exp(-\beta\sigma\omega)\,.\label{A2}
\end{align}
Using the mode expansion of the massless scalar field \cite{peskin2018introduction}
\begin{equation}
\phi(x(\tau))=\frac{1}{(2\pi)^{3/2}}\int_{-\infty}^{+\infty}\frac{d^{3}\mathbf{k}}{\sqrt{2 \omega_{\mathbf{k}}}}\Big[a_{\mathbf{k}}e^{-i\omega_{\mathbf{k}}t+i\mathbf{k}\cdot\mathbf{x}}+a_{\mathbf{k}}^{\dagger}e^{i\omega_{\mathbf{k}} t-i\mathbf{k}\cdot\mathbf{x}}\Big]\,\label{phiS}
\end{equation}
$\langle\sigma\vert\phi(x(\tau))\phi(x(\tau'))\vert \sigma\rangle$ becomes
\begin{align}
&\langle\sigma\vert\phi(x(\tau))\phi(x(\tau'))\vert \sigma\rangle\nonumber\\
=&\frac{1}{(2\pi)^{3}}\int_{-\infty}^{+\infty}\frac{d^{3}\mathbf{k}d^{3}\mathbf{k'}}{\sqrt{4 \omega_{\mathbf{k}}\omega_{\mathbf{k'}}}}\Big[\langle\sigma\vert a_{\mathbf{k}}a_{\mathbf{k'}}^{\dagger}\vert\sigma\rangle e^{-i(\omega_{\mathbf{k}}t-\omega_{\mathbf{k'}}t')+i(\mathbf{k}\cdot\mathbf{x}-\mathbf{k'}\cdot\mathbf{x'})}+\langle\sigma\vert a_{\mathbf{k}}^{\dagger}a_{\mathbf{k'}}\vert\sigma\rangle e^{i(\omega_{\mathbf{k}}t-\omega_{\mathbf{k'}}t')-i(\mathbf{k}\cdot\mathbf{x}-\mathbf{k'}\cdot\mathbf{x'})}\Big]\,.\label{A3}
\end{align}
Now using the relation between $a_{\mathbf{k}}$ and $a_{\mathbf{k}}^{\dagger}$
\begin{equation}
\Big[a_{\mathbf{k}},a_{\mathbf{k}}^{\dagger}\Big]=\delta^{3}(\mathbf{k}-\mathbf{k'})\,.\label{A4}
\end{equation}
Using the above commutation relation eq. \eqref{A3} becomes
\begin{equation}
\langle\sigma\vert\phi(x(\tau))\phi(x(\tau'))\vert \sigma\rangle
=\frac{1}{(2\pi)^{3}}\int_{-\infty}^{+\infty}\frac{d^{3}\mathbf{k}}{2 \omega_{\mathbf{k}}}\Big[(\sigma+1)e^{-i\omega_{\mathbf{k}}(t-t')+i\mathbf{k}\cdot(\mathbf{x}-\mathbf{x'})}+\sigma e^{i\omega_{\mathbf{k}}(t-t')-i\mathbf{k}\cdot(\mathbf{x}-\mathbf{x'})}\Big]\,.\label{A5}
\end{equation}
For the massless scalar field $\omega_{\mathbf{k}}=\vert \mathbf{k}\vert\equiv k$ and defining $t-t'\equiv \Delta t$ and $\mathbf{x}-\mathbf{x'}\equiv \Delta\mathbf{x}$ and putting these values in eq.\eqref{A2}, we get
\begin{align}
&G^{+}_{\beta}(x(\tau),x(\tau'))\nonumber\\
&=\frac{1}{(2\pi)^{3}}\int_{-\infty}^{+\infty}\frac{d^{3}\mathbf{k}}{2 k}\bigg[\displaystyle\sum_{\sigma=0}^{\infty}(\sigma+1)e^{-\beta k\sigma}e^{-i k\Delta t+i\mathbf{k}\cdot\Delta\mathbf{x}}+\displaystyle\sum_{\sigma=1}^{\infty}\sigma e^{-\beta k\sigma}e^{ik\Delta t-i\mathbf{k}\cdot\Delta\mathbf{x}}\bigg]\bigg/\displaystyle\sum_{\sigma=0}^{\infty}e^{-\beta k\sigma}\nonumber\\
&=\frac{1}{(2\pi)^{3}}\int_{-\infty}^{+\infty}\frac{d^{3}\mathbf{k}}{2 k}\bigg[\frac{e^{\beta k}}{e^{\beta k}-1}e^{-i k\Delta t+i\mathbf{k}\cdot\Delta\mathbf{x}}+\frac{1}{e^{\beta k}-1}e^{ik\Delta t-i\mathbf{k}\cdot\Delta\mathbf{x}}\bigg]\nonumber\\
&=\frac{1}{4\pi^2}\left[\frac{1}{\vert \Delta\mathbf{x}\vert}\int_{-\infty}^{+\infty}\frac{\sin(k\vert \Delta\mathbf{x}\vert) e^{i k (\Delta t-i\varepsilon)}}{(e^{\beta k}-1)}dk\right]\hspace{2cm} [\text{where}\,\,\varepsilon\rightarrow 0]\nonumber\\
&=\frac{1}{4\pi^2}\left[\frac{\pi}{2\beta\vert \Delta\mathbf{x}\vert}\left\{\coth\left(\frac{\pi(\vert \Delta\mathbf{x}\vert-\Delta t+i\varepsilon)}{\beta}\right)+\coth\left(\frac{\pi(\vert \Delta\mathbf{x}\vert+\Delta t-i\varepsilon)}{\beta}\right)\right\}\right]\nonumber\\
&=-\frac{1}{4\pi^2}\displaystyle\sum_{n=-\infty}^{\infty}\frac{1}{(t(\tau)-t(\tau')-in\beta-i\varepsilon)^2-(x(\tau)-x(\tau'))^2-(y(\tau)-y(\tau'))^2-(z(\tau)-z(\tau'))^2}\,.\label{A6}
\end{align}

\section{Derivation of the thermal Wightman function inside a cavity}\label{Appendix:CC}

In this Appendix, we provide a complete derivation of the thermal Wightman function of the massless scalar field inside a cavity eq.\eqref{WightCS}.\\
Using the definition of the thermal Wightman function given in eq.\eqref{A1}, we get
\begin{align}
&G^{+}_{\beta}(x(\tau),x(\tau'))=\frac{tr[e^{-\beta \mathscr{H}_{F}}\phi(x(\tau))\phi(x(\tau'))]}{tr[e^{-\beta \mathscr{H}_{F}}]}\nonumber\\
=&tr\left[\exp\left\{-\beta \displaystyle\sum_{\mathbf{k}}\omega_\mathbf{k} a^{\dagger}_\mathbf{k} a_\mathbf{k}\right\}\phi(x(\tau))\phi(x(\tau'))\right]\bigg/tr\left[\exp\left\{-\beta \displaystyle\sum_{\mathbf{k}}\omega_\mathbf{k} a^{\dagger}_\mathbf{k} a_\mathbf{k}\right\}\right]\nonumber\\
=&\bigg[\displaystyle\sum_{\sigma=0}^{\infty}exp(-\beta \sigma\omega)\langle\sigma\vert\phi(x(\tau))\phi(x(\tau'))\vert \sigma\rangle\bigg]\bigg/\displaystyle\sum_{\sigma=0}^{\infty}exp(-\beta\sigma\omega)\,.\label{C1}
\end{align}
In the presence of a single reflecting boundary, the mode function of the massless scalar field operator obeys Dirichlet boundary condition $\phi\vert_{z=z_0}=0$ and takes the form 
\begin{equation}
f(t,x,y,z) \sim \sin (k_{z}z)e^{-i k t}e^{ik_{y}y+ik_{x}x}\label{C2}
\end{equation}
Now, using this mode function and computing the term $\langle\sigma\vert\phi(x(\tau))\phi(x(\tau'))\vert \sigma\rangle$ we get
\begin{align}
&\langle\sigma\vert\phi(x(\tau))\phi(x(\tau'))\vert \sigma\rangle\nonumber\\
&=\frac{1}{(2\pi)^{3}}\int_{-\infty}^{+\infty}\frac{d^{3}\mathbf{k}}{2k}\{\cos[k_z(z-z')]-\cos[k_z(z+z')]\}\Big[(\sigma+1)e^{-ik(t-t')+ik_{x}(x-x')+ik_{y}(y-y')}\Big.\nonumber\\
&\hspace{0.2cm}+\Big.\sigma e^{ik(t-t')-ik_{x}(x-x')-ik_{y}(y-y')}\Big]\,.\label{C3}
\end{align}
Inside the cavity of length $L$, the mode function of the massless scalar field operator obeys Dirichlet boundary condition $\phi\vert_{z=z_0}=\phi\vert_{z=L}0$. Using the mode function given in Ref.\cite{Miyamoto_2019} and computing the term $\langle\sigma\vert\phi(x(\tau))\phi(x(\tau'))\vert \sigma\rangle$ we get
\begin{align}
&\langle\sigma\vert\phi(x(\tau))\phi(x(\tau'))\vert \sigma\rangle\nonumber\\
&=\frac{1}{(2\pi)^{3}}\displaystyle\sum_{n=-\infty}^{\infty}\int_{-\infty}^{+\infty}\frac{d^{3}\mathbf{k}}{2k}\{\cos[k_z(z-z'-2nL)]-\cos[k_z(z+z'-2nL)]\}\Big[(\sigma+1)e^{-ik\Delta t+ik_{x}\Delta x+ik_{y}\Delta y}\Big.\nonumber\\
&\hspace{0.2cm}+\Big.\sigma e^{ik\Delta t-ik_{x}\Delta x-ik_{y}\Delta y}\Big]\,.\label{C4}
\end{align}
Therefore, using eq.\eqref{C4} into the last line of eq.\eqref{C1} and computing the summation over $\sigma$, thermal Wightman function takes the form
\begin{align}
&G^{+}_{\beta}(x(\tau),x(\tau'))\nonumber\\
&=\frac{1}{(2\pi)^{3}}\displaystyle\sum_{n=-\infty}^{\infty}\int_{-\infty}^{+\infty}\frac{d^{3}\mathbf{k}}{2k}\{\cos[k_z(z-z'-2nL)]-\cos[k_z(z+z'-2nL)]\}\Big[\frac{e^{\beta k}}{e^{\beta k}-1} e^{-ik\Delta t+ik_{x}\Delta x+ik_{y}\Delta y}\Big.\nonumber\\
&\hspace{0.2cm}+\Big.\frac{1}{e^{\beta k}-1} e^{ik\Delta t-ik_{x}\Delta x-ik_{y}\Delta y}\Big]\,.\label{C5}
\end{align}
Now, after some algebraic manipulation, and solving the angular integrals, we finally get
\begin{align}
&G^{+}_{\beta}(x(\tau),x(\tau'))\nonumber\\
&=-\frac{1}{4\pi^2}\left[\frac{1}{\vert \Delta\mathbf{x_2}\vert}\int_{-\infty}^{+\infty}\frac{\sin(k\vert \Delta\mathbf{x_2}\vert) e^{i k (\Delta t-i\varepsilon)}}{(e^{\beta k}-1)}dk-\frac{1}{\vert \Delta\mathbf{x_1}\vert}\int_{-\infty}^{+\infty}\frac{\sin(k\vert \Delta\mathbf{x_1}\vert) e^{i k (\Delta t-i\varepsilon)}}{(e^{\beta k}-1)}dk\right]\nonumber\\
&=-\frac{1}{4\pi^2}\left[\frac{\pi}{2\beta\vert \Delta\mathbf{x_2}\vert}\left\{\coth\left(\frac{\pi(\vert \Delta\mathbf{x_2}\vert-\Delta t+i\varepsilon)}{\beta}\right)+\coth\left(\frac{\pi(\vert \Delta\mathbf{x_2}\vert+\Delta t-i\varepsilon)}{\beta}\right)\right\}\right.\nonumber\\
&\hspace{2cm}-\left.\frac{\pi}{2\beta\vert \Delta\mathbf{x_1}\vert}\left\{\coth\left(\frac{\pi(\vert \Delta\mathbf{x_1}\vert-\Delta t+i\varepsilon)}{\beta}\right)+\coth\left(\frac{\pi(\vert \Delta\mathbf{x_1}\vert+\Delta t-i\varepsilon)}{\beta}\right)\right\}\right]\label{C6}
\end{align}
where we consider $\varepsilon$ is very small,
$\vert\Delta\mathbf{x_2}\vert=\sqrt{(\Delta x)^2+(\Delta y)^2+(\Delta z-2nL)^2}$ and $\vert\Delta\mathbf{x_1}\vert=\sqrt{(\Delta x)^2+(\Delta y)^2+(z+z'-2nL)^2}$.\\
This equation can be recast as
\begin{align}
&G^{+}_{\beta}(x(\tau),x(\tau'))\nonumber\\
&=-\frac{1}{4\pi^2}\displaystyle\sum_{m=-\infty}^{\infty}\displaystyle\sum_{n=-\infty}^{\infty}\left[\frac{1}{(t-t'-im\beta-i\varepsilon)^2-(x-x')^2-(y-y')^2-(z-z'-2nL)^2}\right.\nonumber\\
&\hspace{0.5cm}-\left.\frac{1}{(t-t'-in\beta-i\varepsilon)^2-(x-x')^2-(y-y')^2-(z+z'-2nL)^2}\right]
\end{align}
This is the form of the thermal Wightman function inside the cavity which is used in eq.\eqref{WightCS}.
\end{appendices}
\bibliographystyle{JHEP.bst}
\bibliography{Reference}
\end{document}